\documentclass[%
 reprint,
 amsmath,amssymb,
 aps,
prb,
 superscriptaddress
]{revtex4-2}

\usepackage{graphicx}
\usepackage{dcolumn}
\usepackage{bm}
\usepackage{color}
\usepackage[colorlinks=true,linkcolor=blue,urlcolor=blue,citecolor=blue]{hyperref}

\begin{document}

\preprint{APS/123-QED}

\title{Measurement-induced phase transition in free bosons}

\author{Kazuki Yokomizo}
\affiliation{Department of Physics, The University of Tokyo, 7-3-1 Hongo, Bunkyo-ku, Tokyo, 113-0033, Japan}
\author{Yuto Ashida}%
\affiliation{Department of Physics, The University of Tokyo, 7-3-1 Hongo, Bunkyo-ku, Tokyo, 113-0033, Japan}
\affiliation{Institute for Physics of Intelligence, University of Tokyo, 7-3-1 Hongo, Tokyo 113-0033, Japan}




%
\begin{abstract}
The competition between quantum many-particle dynamics and continuous monitoring can lead to measurement-induced phase transitions (MIPTs). So far, MIPTs have been extensively explored in fermionic or spin systems. To examine the possibility of an MIPT in bosonic systems, we study the entanglement structure in continuously monitored free bosons with long-range couplings. When the measurement is local, we find that no MIPTs occur because the substantial entanglement generated by the long-range coupling overcomes the entanglement destruction due to the measurement. In contrast, we show that the nonlocal measurement can efficiently suppress the entanglement generation, leading to an MIPT where the bipartite entanglement entropy exhibits the subvolume-to-area law phase transition as the measurement strength is increased. Our numerical results indicate that the critical point should be described by a certain conformal field theory, while the transition does not belong to a conventional universality class such as Berezinskii-Kosterlitz-Thouless class.
\end{abstract}
\pacs{Valid PACS appear here}
\maketitle
%
%

\section{\label{sec1}Introduction}
Quantum entanglement lies at the heart of quantum physics and quantum technologies~\cite{Horodecki2009,Chitambar2019}. While unitary dynamics can generate entanglement, nonunitary time evolution induced by quantum measurements tends to destroy entanglement. The competition between the generation and destruction of entanglement can give rise to a phase transition at the level of single quantum trajectory, where the dynamics is conditioned on measurement outcomes. Such a measurement-induced phase transition (MIPT) manifests itself as the transition in the size scaling of the bipartite entanglement entropy.

Previous theoretical studies on quantum circuits have demonstrated that the increase in the rate of random projection measurements can lead to the transition of the entanglement scaling from the volume law to the area law~\cite{Skinner2019,Li2019,Tang2020,Zabalo2020,Bao2020,Jian2020,Choi2020,Turkeshi2020,Gullans2020,Lang2020,Gullans2020v2,Lavasani2021,Nahum2021,Block2022,Sharma2022,Fisher2023,Garnet2023,Kelly2023}, and some of the predictions have been tested experimentally~\cite{Noel2022,Koh2023,Hoke2023}. MIPTs have been also explored in quantum many-body systems under continuous quantum measurements, where the entanglement entropy of quantum trajectories has been found to exhibit a volume-to-area law transition as the measurement strength is increased~\cite{Fuji2020,Goto2020,Gopalakrishnan2021,Turkeshi2021,Turkeshi2022,Zerba2023,Russomanno2023,Piccitto2023}.

Nevertheless, the fate of MIPTs in free-particle systems is still under debate. In one-dimensional (1D) free fermions with short-range couplings, an early study~\cite{Cao2019} and recent field-theoretical analysis~\cite{Poboiko2023,Jin2024} have proposed the area law at any nonzero measurement strength, while the signatures of possible critical logarithmic-law phases have been observed in several studies~\cite{Alberton2021,Buchhold2021,Szyniszewski2023}. Meanwhile, in 1D free fermions with long-range couplings, the presence of the subvolume-to-area law phase transition has been predicted~\cite{Minato2022,Muller2022}. So far, however, previous studies have focused on fermionic or spin systems in which the dimension of the Hilbert space is finite. The situation in bosonic systems, where even a local Hilbert space is unbounded, has been much less explored. It is thus timely to ask under what conditions a bosonic MIPT can occur.

\begin{figure}[]
\includegraphics[width=8.5cm]{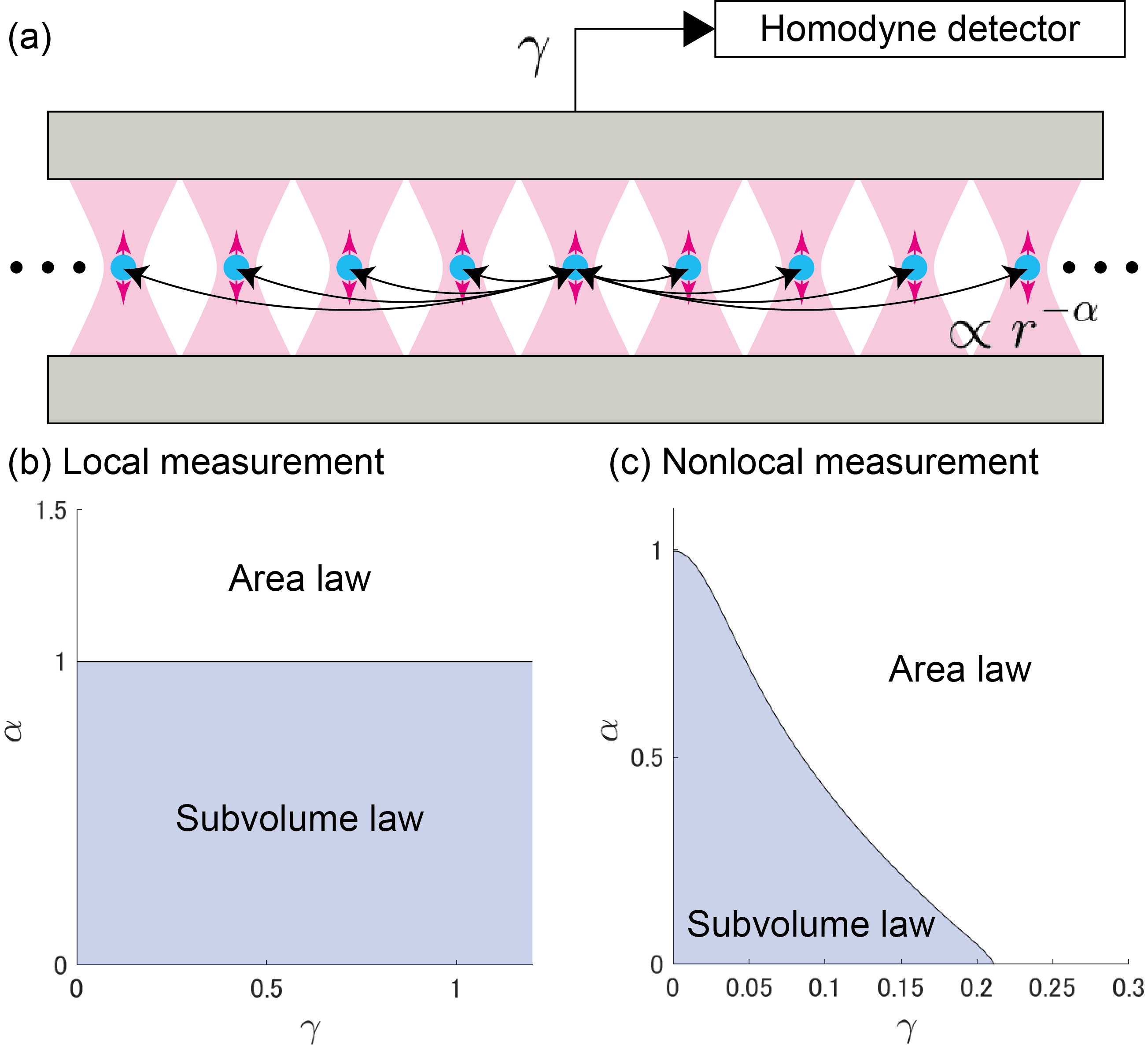}
\caption{\label{fig1}(a) Schematic illustration of the setup. The red arrows represent the direction of the particle motion. The particles interact with each other via the long-range coupling whose strength is proportional to $r^{-\alpha}$ with the distance $r$. We denote the measurement strength by $\gamma$. [(b) and (c)] Phase diagram of the half-chain entanglement entropy $S$, where the subvolume-law phase (respectively area-law phase) exhibits the scaling $S\propto L^b$ with $0<b<1$ (respectively $S\propto L^0$).}
\end{figure}

In this paper, we investigate the entanglement structure in typical individual quantum trajectories of free bosons under continuous monitoring. Specifically, we consider an array of particles [Fig.~\ref{fig1}(a)], each of which is trapped in a harmonic potential. The particles interact with each other via the long-range coupling obeying the power-law decay $r^{-\alpha}$ with the distance $r$ between the particles. The particle positions are continuously monitored via a Gaussian measurement, where a quantum state can be described by a pure bosonic Gaussian state in the long-time limit; this fact allows us to efficiently calculate the bipartite entanglement entropy $S$ for a relatively large system size $L$. Remarkably, we find two distinct types of the phase diagrams [Figs.~\ref{fig1}(b) and \ref{fig1}(c)] depending on whether measurement is short- or long-ranged. Both phase diagrams include the subvolume-law phase with $S\propto L^b~\left(0<b<1\right)$ and the are-law phase with $S\propto L^0$. The main difference between the two cases is that the MIPT, a transition occurring as a function of the measurement strength $\gamma$, can appear only in the case of the nonlocal measurement [Fig.~\ref{fig1}(c)]. The absence of the MIPT under the local measurement [Fig.~\ref{fig1}(b)] is reminiscent of the previous finding in free fermions~\cite{Muller2022}. When taking the short-range limit $\alpha\rightarrow\infty$ in the present setup, our analysis reproduces a recent result~\cite{Minoguchi2022,Young2024} where the area law with no phase transitions has been found. Altogether, our results indicate that the long-rangedness in both particle couplings and measurements is crucial for the MIPT to occur in free-boson systems.

We further reveal that the critical point should be described by a certain conformal field theory (CFT). Meanwhile, the phase transition does not belong to a conventional universality class known in the existing studies on free-fermion systems, such as Refs.~\cite{Alberton2021,Minato2022,Muller2022}, where the numerical results indicate the Berezinskii-Kosterlitz-Thouless class. These results indicate that the model considered in our work, which possesses a large mass gap, goes beyond the Tomonaga-Luttinger liquid-type theory that were explored in, e.g., Refs.~\cite{Buchhold2021,Muller2022}.

Before getting into details, we mention that our study is partly motivated by recent advances in the control of levitated nanoparticles~\cite{Millen2020,Gonzalez2021}. Indeed, it has been experimentally demonstrated that multiple levitated nanoparticles can be prepared in a tunable manner, and they exhibit the long-range interaction with the power-law decay~\cite{Rieser2022}. Furthermore, it has been proposed that continuous observation of the dynamics of trapped particles can be realized by measuring the scattered light via homodyne detection~\cite{Rudolph2022,Rudolph2024}. Our results thus show that nontrivial measurement-induced phenomena can occur in such continuous-variable quantum systems under quantum measurements.

The rest of this paper is organized as follows. We first introduce a framework to explore the entanglement entropy in free-boson systems in Sec.~\ref{sec2}. We next discuss the entanglement transition under the local or nonlocal measurement in Sec.~\ref{sec3}. We further perform a finite-size scaling analysis to investigate a universality class of the entanglement transition in Sec.~\ref{sec4}. Finally, in Sec.~\ref{sec5}, we summarize the results and comment on the perspective of this paper.

%
%

\section{\label{sec2}Model}
To be concrete, we consider the array consisting of $L$ particles that are placed periodically and subject to an open boundary condition. The dynamics of the particles is described by a set of the position and momentum operators denoted by $\hat{\bm\phi}=(\hat{x}_1,\dots,\hat{x}_L,\hat{p}_1,\dots,\hat{p}_L)^{\rm T}$. The total Hamiltonian reads
\begin{equation}
\hat{H}=\sum_{j=1}^L\frac{\Omega}{2}(\hat{p}_j^2+\hat{x}_j^2)+\sum_{r=1}^{L-1}\sum_{j=1}^{L-r}\frac{K}{2r^\alpha}(\hat{x}_j-\hat{x}_{j+r})^2,
\label{eq1}
\end{equation}
where $\Omega$ and $K$ are the trap frequency and the coupling strength, respectively, and $\alpha$ is the exponent that characterizes the power-law decay of the long-range coupling. We note that the short-range coupling corresponds to taking the limit $\alpha\rightarrow\infty$. For the sake of simplicity, we assume $\Omega=K=1$ throughout this paper.

The particles are subject to a continuous Gaussian measurement described by a Hermitian jump operator $\hat{\cal O}_n$, which is given by a linear combination of $\hat{\phi}_j$. In the diffusive limit~\cite{Gisin1984,Diosi1988,Diosi1988v2,Jacobs2006,Ashida2020}, the time evolution of a quantum many-body state $\left|\psi\right\rangle$ is governed by the stochastic Schr\"{o}dinger equation,
\begin{equation}
d\left|\psi\right\rangle=\left[-i\hat{H}dt+\sum_n\left(-\frac{1}{2}\hat{\cal M}_n^2dt+\hat{\cal M}_ndW_n\right)\right]\left|\psi\right\rangle,
\label{eq2}
\end{equation}
where $\hat{\cal M}_n=\hat{\cal O}_n-\langle\hat{\cal O}_n\rangle$, $dW_n$ is the Wiener increment satisfying ${\mathbb E}\left[dW_n\right]=0$ and $dW_ndW_m=\delta_{nm}dt$, and we set $\hbar=1$. Here, $\left\langle\cdots\right\rangle$ is the expectation value with respect to $\left|\psi\right\rangle$, and ${\mathbb E}\left[\dots\right]$ represents the ensemble average over the measurement outcomes. We review a bosonic Gaussian state, which can be fully characterized by its mean value and covariance matrix~\cite{Weedbrook2012}, and the time-evolution under the Gaussian measurement in Appendix~\ref{secA} and Appendix~\ref{secB}, respectively. We further show that, in the long-time limit, an arbitrary initial state under the Gaussian measurement converges to a bosonic Gaussian state in Appendix~\ref{secC}.

In our work, we consider the following two measurement processes. First, the local measurement is defined by the jump operator,
\begin{equation}
\hat{\cal O}_n=\sqrt{\gamma}\hat{x}_n,~\left(n=1,\dots,L\right),
\label{eq3}
\end{equation}
which acts on each particle and includes the parameter $\gamma\left(>0\right)$ as the measurement strength. Second, the nonlocal measurement is represented by
\begin{equation}
\hat{\cal O}_n=\sqrt{\frac{\gamma}{r^\alpha}}(\hat{x}_j\pm\hat{x}_{j+r}).
\label{eq4}
\end{equation}
Here, $n=\left(j,r,\pm\right)$ now includes a set of the indices, where $j$ and $r$ define the particle positions that the jump operator acts on, and $\pm$ characterizes whether the jump operator is a symmetric or antisymmetric combination of the position operators. The nonlocality here refers to the fact that the jump operator acts on the two distant particles.

A key quantity to analyze the entanglement structure of the typical quantum trajectory is the ensemble-averaged entanglement entropy in the steady-state regime. Specifically, we divide the whole system into two regions $A$ and $\bar{A}$ and introduce the reduced density operator $\hat{\rho}_A={\rm Tr}_{\bar{A}}(\left|\psi\right\rangle\left\langle\psi\right|)$ for each quantum trajectory. We then define the typical entanglement entropy $S_A$ by taking the ensemble average of the von Neumann entropy over the measurement outcomes: $S_A=-{\mathbb E}[{\rm Tr}(\hat{\rho}_A\ln\hat{\rho}_A)]$. Importantly, the analysis of $S_A$ can be greatly simplified in the present setup because the time evolution of the covariance matrix turns out to be deterministic and has a unique steady solution~\cite{Minoguchi2022}. Consequently, the ensemble average is not necessary, but it suffices to analyze the covariance matrix in the steady-state regime.

Below, we focus on the half-chain entanglement entropy, which can be calculated from the symplectic eigenvalues of the submatrix of the covariance matrix corresponding to the subregion $A=\left\{j|1\leq j\leq L/2\right\}$. We also analyze the mutual information $I_{BC}=S_B+S_C-S_{B\cup C}$ between distant subregions $B=\left\{j|L/4+1\leq j\leq 3L/8\right\}$ and $C=\left\{j|5L/8+1\leq j\leq 3L/4\right\}$, which can be used as yet another indicator of a MIPT.

%
%

\section{\label{sec3}Local or nonlocal measurement}
\begin{figure}[]
\includegraphics[width=8.5cm]{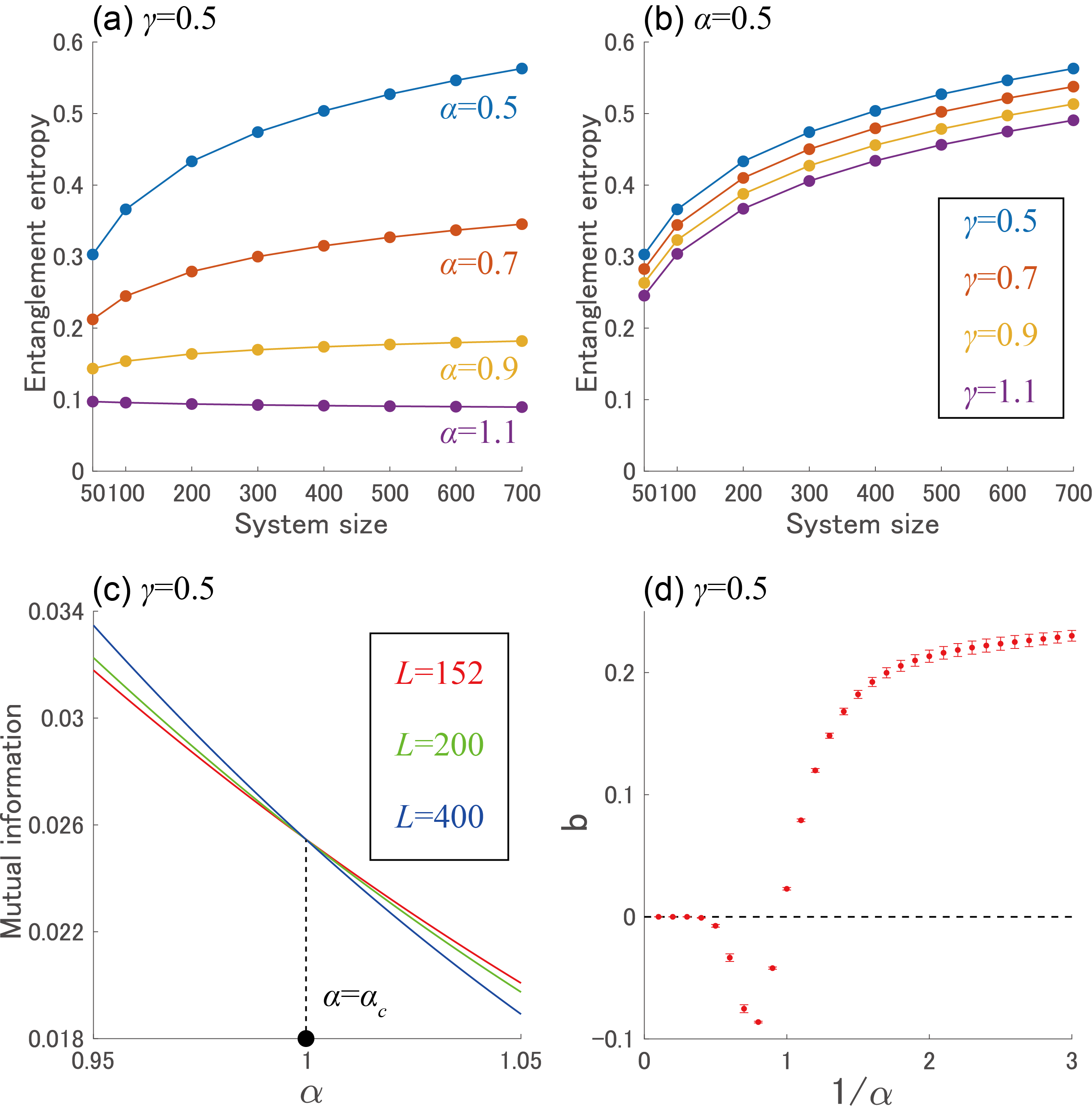}
\caption{\label{fig2}Finite-size analysis under the local measurement. [(a) and (b)] Half-chain entanglement entropy. (c) Mutual information $I_{BC}$. (d) Size-scaling exponent $b$ extracted from the fitting function, $S_A=aL^b$. The error bars indicate a $2\sigma$ confidential interval.}
\end{figure}
\begin{figure}[b]
\includegraphics[width=8.5cm]{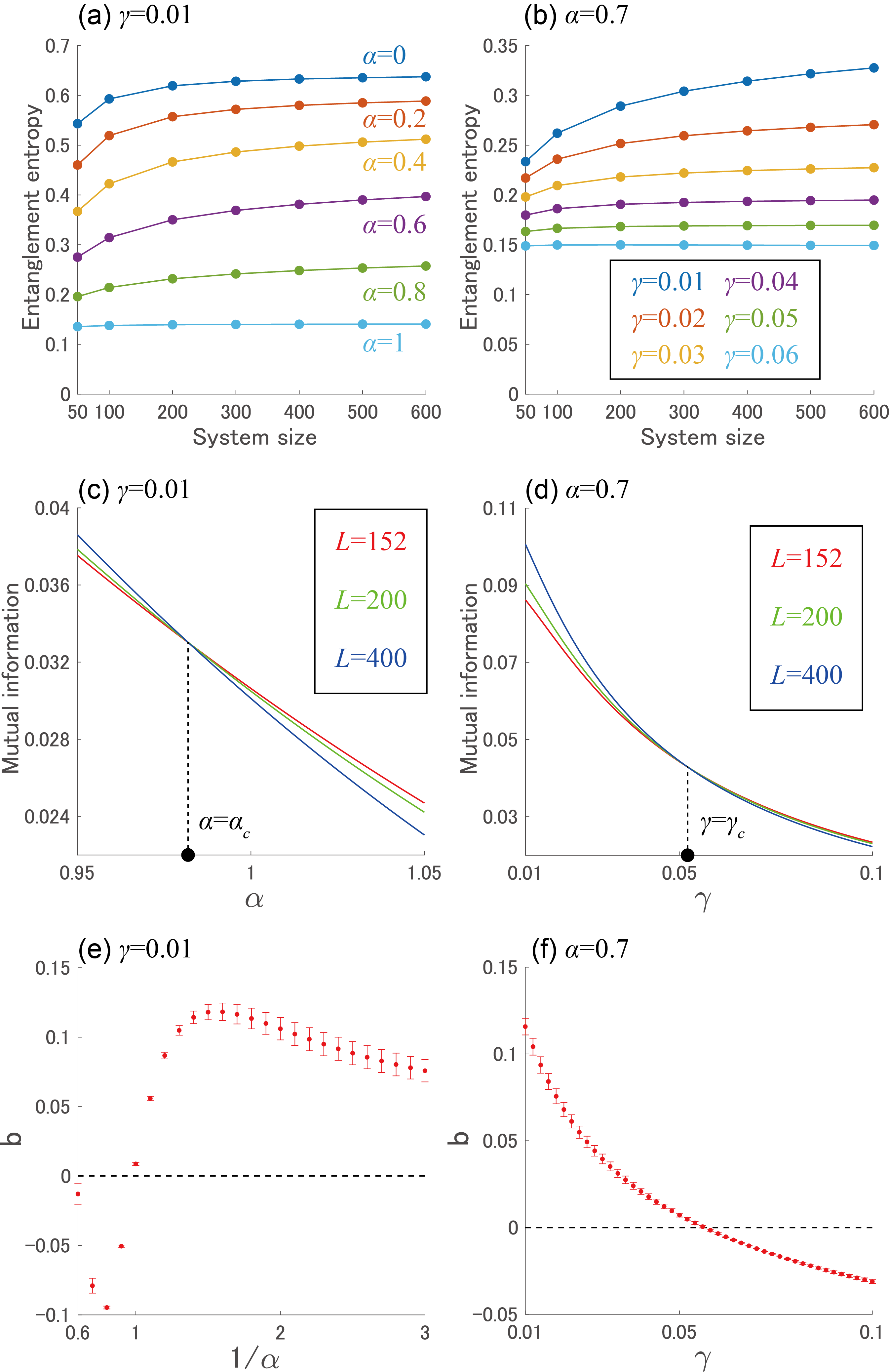}
\caption{\label{fig3}Finite-size analysis under the nonlocal measurement. [(a) and (b)] Half-chain entanglement entropy. [(c) and (d)] Mutual information $I_{BC}$. [(e) and (f)] Size-scaling exponent $b$ extracted from the fitting function, $S_A=aL^b$. The error bars indicate a $2\sigma$ confidential interval.}
\end{figure}
We first consider the case of the local measurement given in Eq.~(\ref{eq3}). In this case, on the one hand, it has been known that no MIPTs are expected in the short-range limit, $\alpha\rightarrow\infty$, because the correlation function decays exponentially, and the entanglement entropy obeys the area law at any measurement strength~\cite{Minoguchi2022,Young2024}. On the other hand, however, the long-range coupling can efficiently transfer the quantum information between the distant particles and should facilitate the entanglement generation. As $\alpha$ is decreased, the enhanced entanglement generation is expected to eventually overcome the entanglement destruction due to the measurement. As a result, the entanglement phase transition between the subvolume- and area-law phases occurs at $\alpha=1$ independent of the measurement strength $\gamma$ [Figs.~\ref{fig2}(a) and \ref{fig2}(b)]. Indeed, we find that the entanglement scaling changes from the area law to the subvolume law, from the numerical data of the mutual information intersecting at $\alpha_c=1$ [Fig.~\ref{fig2}(c)]. Accordingly, spatial decay of the correlation function obeys an exponential law when $\alpha>\alpha_c$ or a power law when $\alpha<\alpha_c$ (cf. Appendix~\ref{secD}); the latter indicates that the subvolume-law phase exhibits the critical behavior. In the subvolume-law phase at $\alpha<\alpha_c$, the entanglement entropy obeys the algebraic scaling, $S_A\propto L^b$ with $0<b<1$. Figure~\ref{fig2}(d) shows the dependence of the size-scaling exponent $b$ on $1/\alpha$. The point where the sign of $b$ changes from positive to negative corresponds to the transition point, which agrees with the critical point extracted from Fig.~\ref{fig2}(c).

Physically, these findings can be understood from growth rate of the entanglement entropy in bosonic systems, which is characterized by the Lyapunov exponent of the Hamiltonian in subregion $A$~\cite{Bianchi2018,Hackl2018}: namely, the growth rate diverges in the thermodynamic limit at arbitrary measurement strength when $\alpha<1$, and the local measurement is simply not enough to suppress such substantial entanglement growth (cf. Appendix~\ref{secE}). This indicates that the transition point is independent of the measurement strength, leading to the phase diagram [Fig.~\ref{fig1}(b)]. All in all, under the local measurement, the system can exhibit the entanglement transition which essentially originates from the diverging growth rate of the entanglement generated by the long-range coupling rather than the nontrivial competition with the measurement. In this respect, we argue that the free bosons under the local measurement do not show a MIPT that should occur as a function of the measurement strength.

To explore the possibility of MIPTs in bosonic systems, we next consider the system under the nonlocal measurement given in Eq.~(\ref{eq4}). Because of the nonlocal nature of the jump operator, we expect that the measurement can now suppress the rapid entanglement growth generated by the long-range coupling more efficiently. As a result, the entanglement phase transition depends on both $\alpha$ and $\gamma$ [Figs.~\ref{fig3}(a) and \ref{fig3}(b)], which indicate that the measurement-induced phase transition as a function of the measurement strength occurs only under the nonlocal measurement.

By varying $\alpha$, we find the intersecting point indicated by $\alpha=\alpha_c$ in Fig.~\ref{fig3}(c) and confirm that the critical point agrees with the one extracted from the size-scaling exponent $b$ [Fig.~\ref{fig3}(e)]. We remark that, in Fig.~\ref{fig3}(e), there is an optimal $\alpha$ at which the size-scaling exponent becomes maximal. This nonmonotonic behavior is absent in the case of the local measurement [cf. Fig.~\ref{fig2}(c)] and should have its roots in the fact that the nonlocal measurement tends to overcome the entanglement generation in the infinite-range limit $1/\alpha\rightarrow\infty$.

We further find that the system undergoes the subvolume-to-area law phase transition as the measurement strength $\gamma$ is increased [Fig.~\ref{fig3}(d)], making a sharp contrast to the case of the local measurement. Accordingly, the phase boundary now depends on $\gamma$ in addition to $\alpha$ as shown in Fig.~\ref{fig1}(c), where we numerically determine the transition point by locating the intersecting point in the mutual information. In the subvolume-law phase at $\gamma<\gamma_c$, the half-chain entanglement entropy obeys $S_A\propto L^b$ with $0<b<1$, similar to the case of the local measurement. Figure~\ref{fig3}(e) plots the size-scaling exponent $b$ as a function of $\gamma$, and the critical point again agree with the one extracted from Fig.~\ref{fig3}(d).

We briefly discuss the case of the short-ranged free bosons under the nonlocal measurement described by Eq.~(\ref{eq4});~\footnote{We note that the exponent $\alpha$ here only refers to the long-distance decay in the jump operator in Eq.~(\ref{eq4})} a similar setup has recently been considered in Refs.~\cite{Russomanno2023,Albornoz2024} for free fermions. Our result shows no evidences of entanglement transitions, where the entanglement entropy always obeys the area law (cf. Appendix~\ref{secF}).

%
%

\section{\label{sec4}Finite-size scaling analysis}
\begin{figure}[]
\includegraphics[width=8.5cm]{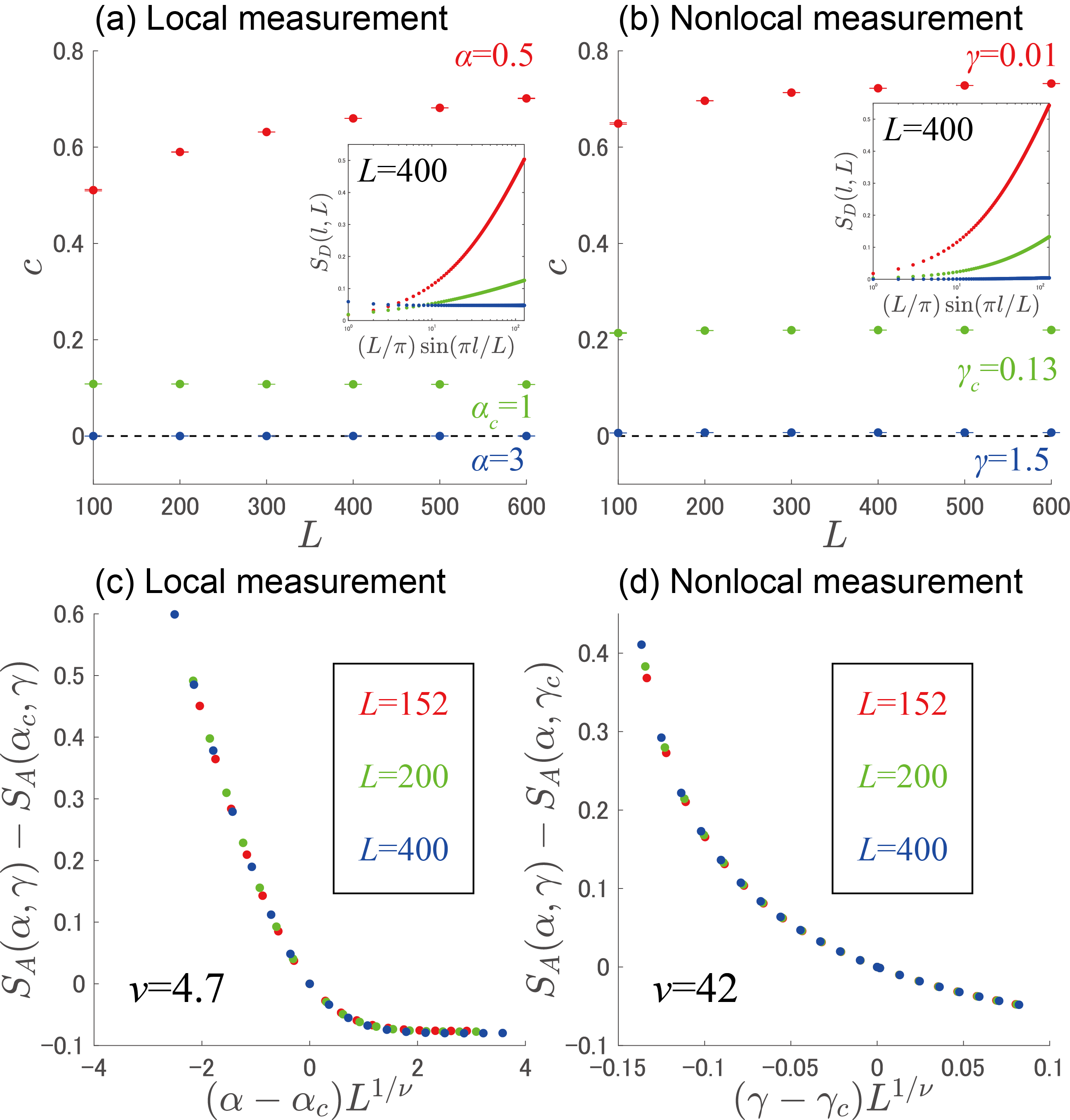}
\caption{\label{fig4}Finite-size scaling analysis under the [(a) and (c)] local or [(b) and (d)] nonlocal measurement. [(a) and (b)] Estimated value of the central charge in the subvolume-law phase (red), at the transition point (green), and in the area-law phase (blue). The insets plot the entanglement entropy as a function of the chord length in a semilogarithmic plot. [(c) and (d)] Data collapse for the half-chain entanglement entropy. The scaling functions are defined in Eqs.~(\ref{eq6}) and (\ref{eq7}) in (c) and (d), respectively. In (d), the entanglement entropy at the critical point is identical across all the system sizes to two significant digits, with a value of $S_A\left(\alpha,\gamma_c\right)=0.13$. We set $\gamma=0.5$ in [(a) and (b)] or $\alpha=0.3$ in [(b) and (d)].}
\end{figure}

Finally, we investigate a possible characterization by a CFT, which predicts that the entanglement entropy for the subregion $D=\left\{j|1\leq j\leq l\right\}$ obeys
\begin{equation}
S_D=\frac{c}{3}\ln\left(\frac{L}{\pi}\sin\frac{\pi l}{L}\right)+s_0,
\label{eq5}
\end{equation}
where $c$ and $s_0$ represent the central charge and the residual entropy, respectively. Figures.~\ref{fig4}(a) and \ref{fig4}(b) plot the system-size dependence of the effective central charge extracted by fitting the numerical results to Eq.~(\ref{eq5}); the insets show the entanglement entropy plotted as a function of the chord length. We find that, in the subvolume-law phase (red), the estimated central charge depends on the system size and monotonically increases, while it almost vanishes in the area-law phase (blue). We remark that the dependence of the central charge on the system size indicates the power-law growth of the entanglement entropy~\cite{Millen2020}. We also find that, at the transition point (green), the central charge takes a nonzero constant value independent of the system size. These results indicate that the critical behavior at the transition point should be described by a certain CFT.

We next perform a finite-size scaling analysis to investigate the universality class of the phase transition. We assume that, in the vicinity of the transition point, the half-chain entanglement entropy obeys
\begin{equation}
S_A\left(\alpha,\gamma\right)-S_A\left(\alpha_c,\gamma\right)=F\left[\left(\alpha-\alpha_c\right)L^{1/\nu}\right]
\label{eq6}
\end{equation}
in the case of the local measurement or
\begin{equation}
S_A\left(\alpha,\gamma\right)-S_A\left(\alpha,\gamma_c\right)=F\left[\left(\gamma-\gamma_c\right)L^{1/\nu}\right]
\label{eq7}
\end{equation}
in the case of the nonlocal measurement, where $F\left(x\right)$ is a smooth function, and $\nu$ is a critical exponent. Figures~\ref{fig4}(c) and \ref{fig4}(d) show the data collapse of the entanglement entropy into a single curve. We extract the critical exponent $\nu=4.7\pm0.5$ in the case of the local measurement or $\nu=42\pm2$ in the case of the nonlocal measurement (cf. Appendix~\ref{secG}). The values of $c$ and $\nu$ estimated from the data collapses at least do not allow for a straightforward CFT explanation based on, e.g., minimal models or the Tomonaga-Luttinger liquid. The results also suggest that the phase transitions should not belong to the Berezinskii-Kosterlitz-Thouless class.

%
%

\section{\label{sec5}Discussion and Summary}
We briefly discuss possible experimental relevance of our setup illustrated in Fig.~\ref{fig1}(a). The array of continuous-variable quantum systems considered here would be prepared by periodically arranging multiple levitated nanoparticles. Indeed, all the particles are coupled by the long-range interaction known as the optical binding force~\cite{Rieser2022}. Measuring the light scattered from the particles by, e.g., homodyne detection, one might implement the continuous monitoring of the particle positions ~\cite{Rudolph2022,Rudolph2024}. The stochastic quantum dynamics of the monitored levitated nanoparticles can be then described by Eq.~(\ref{eq2}) in the ideal limit of the perfect detection efficiency. For instance, the local jump operator corresponds to the position measurement of each particle, while the nonlocal one could be realized by conducting the basis transformation in the homodyne detection of the two distant particles~\cite{Rudolph2022,Rudolph2024}. It is worth noting that, even under imperfect measurements, the half-chain logarithmic negativity also exhibits the subvolume-to-are law phase transition (cf. Appendix~\ref{secH}), indicating that the mixed state can also show the entanglement transition.

In summary, we examine under what conditions the MIPT can occur in the free-boson systems under continuous monitoring. Our key finding is that the MIPT, a transition induced by increasing the measurement strength, can occur only when both particle-particle couplings and measurements are long-ranged. The results are summarized in the phase diagrams in Figs.~\ref{fig1}(b) and \ref{fig1}(c); under the local measurement, the phase boundary between the subvolume-law and area-law phases is insensitive to the measurement strength $\gamma$, while the boundary acquires the $\gamma$ dependence in the case of the nonlocal measurement, thus leading to the MIPT. Finally, we reveal that the critical point should be described by a certain CFT, and the transition should belong to an unconventional universality class other than, e.g., the Berezinskii-Kosterlitz-Thouless class.

Several interesting directions remain for future studies. First, it should be intriguing to study the universality class of the MIPT from a field-theoretical perspective if at all possible. In general, it is highly nontrivial to construct a faithful effective theory to describe a phase transition in the so-called strong long-range coupling regime, characterized by the exponent $0\leq\alpha\leq1$, as considered in the present work. Indeed, in this regime, conventional statistical mechanics can be invalid due to the nonadditivity~\cite{Defenu2023}, and it has been recently reported that the Tomonaga-Luttinger liquid description fails to describe the low-energy behavior of strongly long-range interacting bosonic systems~\cite{Botzung2021}. We recall that our model (\ref{eq1}) includes the one-particle harmonic potential leading to the large energy gap, which also makes it nontrivial to apply an effective theory obtained by perturbing the Tomonaga-Luttinger liquid, such as the sine-Gordon theory. The critical behavior in the subvolume-law phase also implies that a description beyond a CFT might be necessary (cf. Appendix~\ref{secD}).

Second, we note that the Lindblad evolution of the present setup, which corresponds to the dynamics obtained after taking the ensemble average, correctly reproduces the equations of motion for the dissipative dynamics of levitated nanoparticle arrays that have been originally considered in the classical regime~\cite{Yokomizo2023}. There, one can implement the nonreciprocal dipole-dipole interaction as experimentally realized in the case of two levitated nanoparticles~\cite{Rieser2022}. It would be interesting to explore the role of nonreciprocal interaction in a full-fledged quantum trajectory in monitored  many-particle systems beyond no-click limit~\cite{Kawabata2023,Lee2024}.

Finally, it is also natural to investigate the entanglement structure in the ensemble-averaged density operator rather than the one in the typical quantum trajectory considered here. Our initial analysis shows that the half-chain logarithmic negativity vanishes in short time (cf. Appendix~\ref{secH}). While the vanishment of the logarithmic negativity does not imply the separability in general, necessary and sufficient criteria for the separability of a bosonic Gaussian state has been known~\cite{Werner2001,Plenio2005,Lami2018}. It merits further study to understand the entanglement structure of levitated nanoparticle arrays under decoherence from the viewpoint of those criteria.

%
%

\section*{\label{DA}Data availability}
The datasets obtained in our work are available in the GitHub repository at \url{https://github.com/Yokomizo0226/MIPTFB-dataset.git}.

%
%

\begin{acknowledgments}
We are grateful to Yohei Fuji, Keisuke Fujii, Zongping Gong, Masahiro Hoshino, Keiji Saito, and Benjamin A. Stickler for variable discussions. K.Y. is supported by JSPS KAKENHI through Grant No.~JP21J01409. Y.A. acknowledges support from the Japan Society for the Promotion of Science through Grant No.~JP19K23424 and from JST FOREST Program (Grant No.~JPMJFR222U, Japan).
\end{acknowledgments}

%
%

\appendix

%
%

\section{\label{secA}Bosonic Gaussian state and entanglement entropy}
In this section, we define a bosonic Gaussian state and summarize its key results (see Ref.~\cite{Weedbrook2012} for further details). We also explain how one can calculate the von Neumann entropy for a general bosonic Gaussian state, which measures the degree of quantum entanglement between a subsystem and the rest of the system. Suppose that $N$ bosonic modes are described by a vector of the linear operators:
\begin{equation}
\hat{\bm\phi}=(\hat{x}_1,\dots,\hat{x}_N,\hat{p}_1,\dots,\hat{p}_N)^{\rm T},
\label{eqappa1}
\end{equation}
which satisfy the canonical commutation relations,
\begin{eqnarray}
[\hat{\phi}_j,\hat{\phi}_k]=i\sigma_{jk},~\sigma=\left( \begin{array}{cc}
O    & 1_N \vspace{3pt}\\
-1_N & 0
\end{array}\right),
\label{eqappa2}
\end{eqnarray}
where $1_N$ is an $N\times N$ identity matrix, and we set $\hbar=1$. Let $\hat{\rho}$ denote a density operator of an $N$-mode bosonic many-body state. The density operator can be represented as a quasiprobability distribution on the phase space in general. We define the displacement operator as
\begin{equation}
\hat{D}_{\bm\xi}=\exp(i\hat{\bm\phi}^{\rm T}\sigma{\bm\xi}),
\label{eqappa3}
\end{equation}
where ${\bm\xi}\in{\mathbb R}^{2N}$. The definition of the characteristic function then reads
\begin{equation}
\chi\left({\bm\xi}\right)={\rm Tr}(\hat{\rho}\hat{D}_{\bm\xi}),
\label{eqappa4}
\end{equation}
and the Wigner function is defined as
\begin{equation}
W\left({\bm\zeta}\right)=\int_{{\mathbb R}^{2N}}\frac{d^{2N}{\bm\xi}}{\left(2\pi\right)^{2N}}\exp\left(-i{\bm\zeta}^{\rm T}\sigma{\bm\xi}\right)\chi\left({\bm\xi}\right),
\label{eqappa5}
\end{equation}
where ${\bm\zeta}\in{\mathbb R}^{2N}$. We here remark that the statistical moments of $\hat{\bm\phi}$ characterize the above phase-space representation. In particular, the defining features in the bosonic Gaussian state are the first moment defined as the mean value,
\begin{equation}
{\bm\phi}=\langle\hat{\bm\phi}\rangle={\rm Tr}(\hat{\rho}\hat{\bm\phi}),
\label{eqappa6}
\end{equation}
and the second moment defined as the covariance matrix,
\begin{equation}
\left(\Gamma_\phi\right)_{jk}=\frac{1}{2}\langle\{\delta\hat{\phi}_j,\delta\hat{\phi}_k\}\rangle
\label{eqappa7}
\end{equation}
for $j,k=1,\dots,2N$, where $\delta\hat{\phi}_j=\hat{\phi}_j-\langle\hat{\phi}_j\rangle$. We note that the covariance matrix satisfies the Heisenberg uncertainty relation given by
\begin{equation}
\Gamma_\phi+\frac{i}{2}\sigma\geq0,
\label{eqappa8}
\end{equation}
where the equality holds when the quantum state is a pure state. A bosonic Gaussian state is fully characterized by the first and second moments. More specifically, the many-body state whose Wigner function can be written as the Gaussian form,
\begin{equation}
W\left({\bm\zeta}\right)=\frac{1}{\left(2\pi\right)^N\sqrt{\det\Gamma_\phi}}\exp\left[\frac{1}{2}\left({\bm\zeta}-{\bm\phi}\right)^{\rm T}\Gamma_\phi^{-1}\left({\bm\zeta}-{\bm\phi}\right)\right],
\label{eqappa9}
\end{equation}
is classified as the bosonic Gaussian state. One of the key properties of the Gaussian states is that any moments of $\hat{\bm\phi}$ can be calculated from the first and second moments, which is nothing but Wick's theorem. Let us consider the $n$-point correlation function defined by
\begin{equation}
C_n=\langle(\hat{\phi}_{j_1}-\phi_{j_1})\cdots(\hat{\phi}_{j_n}-\phi_{j_n})\rangle.
\label{eqappa10}
\end{equation}
For a Gaussian state, any odd-ordered $n$-point correlation functions vanish, i.e.,
\begin{equation}
C_{2m+1}=0,~\left(m\in{\mathbb N}\right),
\label{eqappa11}
\end{equation}
while even-ordered $n$-point correlation functions can be represented by products of the two-point correlation functions~\cite{Hackl2018,Hackl2021}.

We next derive the formula of the entanglement entropy for a bosonic Gaussian state. We consider the bipartition of the system into two subsystems $A$ and $\bar{A}$, where $A$ contains the $N_A$ bosonic modes. Suppose that $\hat{\rho}_A={\rm Tr}_{\bar{A}}(\hat{\rho_{\rm G}})$ represents the reduced density operator of the subsystem $A$, where $\hat{\rho}_{\rm G}$ is the density operator of the Gaussian state. We here focus on the von Neumann entropy between the subsystems $A$ and $\bar{A}$, defined as
\begin{equation}
S_A=-{\rm Tr}(\hat{\rho}_A\ln\hat{\rho}_A).
\label{eqappa12}
\end{equation}
It is worth noting that Eq.~(\ref{eqappa12}) only depends on the covariance matrix $\left(\Gamma_\phi\right)_A$ obtained from $\hat{\rho}_A$, since it is possible to arbitrarily adjust the first moment to be zero by local unitary operations. To calculate Eq.~(\ref{eqappa12}), we recall Williamson's theorem~\cite{Arnold1989}, which states that there exists a symplectic matrix $\bar{S}$ diagonalizing the covariance matrix in the form of
\begin{equation}
\bar{S}\left(\Gamma_\phi\right)_A\bar{S}^{\rm T}={\rm diag}\left(\kappa_1,\dots,\kappa_{N_A},\kappa_1,\dots,\kappa_{N_A}\right),
\label{eqappa13}
\end{equation}
where $\kappa_l~\left(l=1,\dots,N_A\right)$ are called symplectic eigenvalues. We note that all the symplectic eigenvalues take values greater than or equal to $1/2$ because of the Heisenberg uncertainty relation. Importantly, the matrix $\bar{S}$ generates the unitary operator $\hat{U}$ transforming $\hat{\rho}_A$ into the factorized form given by
\begin{eqnarray}
\hat{\rho}_A&=&\hat{U}\hat{\rho}^\otimes\hat{U}^\dag, \label{eqappa14}\\
\hat{\rho}^\otimes&=&\bigotimes_{l=1}^{N_A}\hat{\rho}_l, \label{eqappa15}\\
\hat{\rho}_l&=&\frac{1}{\kappa_l+1/2}\sum_{n=0}^\infty\left(\frac{\kappa_l-1/2}{\kappa_l+1/2}\right)^n\left|n\right\rangle\left\langle n\right|, \label{eqappa16}
\end{eqnarray}
where $\left|n\right\rangle$ is the $n$th Fock state~\cite{Adesso2004}. Thus, the above representation allows us to calculate Eq.~(\ref{eqappa12}) as follows:
\begin{eqnarray}
S_A&=&-\sum_{l=1}^{N_A}{\rm Tr}(\hat{\rho}_l\ln\hat{\rho}_l) \nonumber\\
&=&\sum_{l=1}^{N_A}\left[\left(\kappa_l+\frac{1}{2}\right)\ln\left(\kappa_l+\frac{1}{2}\right)\right. \nonumber\\
&&\left.-\left(\kappa_l-\frac{1}{2}\right)\ln\left(\kappa_l-\frac{1}{2}\right)\right].
\label{eqappa17}
\end{eqnarray}

%
%

\section{\label{secB}Time-evolution equation}
In this section, we derive the time-evolution equation of the bosonic Gaussian state under continuous monitoring (e.g., Ref.~\cite{Minoguchi2022}). We consider an $N$ mode bosonic Gaussian state $\left|\psi_{\rm G}\right\rangle$. The dynamics of $\left|\psi_{\rm G}\right\rangle$ under continuous monitoring can be described by the following stochastic Sch\"{o}dinger equation:
\begin{eqnarray}
&&d\left|\psi_{\rm G}\right\rangle=\left[-i\hat{H}dt-\frac{1}{2}\sum_n(\hat{\cal O}_n-\langle\hat{\cal O}_n\rangle)^2dt\right. \nonumber\\
&&\left.+\sum_n(\hat{\cal O}_n-\langle\hat{\cal O}_n\rangle)dW_n\right]\left|\psi_{\rm G}\right\rangle,
\label{eqappb1}
\end{eqnarray}
where $\hat{H}$ and $\hat{\cal O}_n$ represent a Hamiltonian and measurement operators, respectively, $dW_n$ is the Wiener increment which has the zero mean, ${\mathbb E}\left(dW_n\right)=0$, and satisfies $dW_ndW_{n^\prime}=\delta_{nn^\prime}dt$, and $\left\langle\cdots\right\rangle=\left\langle\psi_{\rm G}\right|\cdots\left|\psi_{\rm G}\right\rangle$. We assume that the Hamiltonian can be written by the quadratic form,
\begin{equation}
\hat{H}=\frac{1}{2}\sum_{j,k}h_{jk}\hat{\phi}_j\hat{\phi}_k,
\label{eqappb2}
\end{equation}
with the symmetric matrix $h$ and that the measurement operators are the linear combinations of $\hat{\bm\phi}$,
\begin{equation}
\hat{\cal O}_n=\sum_jO_{nj}\hat{\phi}_j,
\label{eqappb3}
\end{equation}
with the real-valued matrix $O$. Under the Gaussian measurement, arbitrary initial states converge to bosonic Gaussian states in the long-time limit, as mentioned below. Thus we consider only the time evolutions of the mean value,
\begin{equation}
{\bm\phi}=\left\langle\psi_{\rm G}\right|\hat{\bm\phi}\left|\psi_{\rm G}\right\rangle,
\label{eqappb4}
\end{equation}
and the covariance matrix,
\begin{equation}
\left(\Gamma_\phi\right)_{jk}=\frac{1}{2}\left\langle\psi_{\rm G}\right|\{\delta\hat{\phi}_j,\delta\hat{\phi}_k\}\left|\psi_{\rm G}\right\rangle
\label{eqappb5}
\end{equation}
for $j,k=1,\dots,2N$, where $\delta\hat{\phi}_j=\hat{\phi}_j-\langle\hat{\phi}_j\rangle$.

To derive the time-evolution equation of the first-moment of the Gaussian state, we note that, for an arbitrary operator $\hat{M}$, one can derive the following equation from Eq.~(\ref{eqappb1}):
\begin{eqnarray}
&&d\langle\hat{M}\rangle=-\left(i\langle[\hat{M},\hat{H}]\rangle+\frac{1}{2}\sum_n\langle[\hat{\cal O}_n,[\hat{\cal O}_n,\hat{M}]]\rangle\right)dt \nonumber\\
&&+\sum_n\langle\{\hat{\cal O}_n,\hat{M}\}-2\langle\hat{\cal O}_n\rangle\hat{M}\rangle dW_n.
\label{eqappb6}
\end{eqnarray}
Since we have the relations,
\begin{eqnarray}
-i[\hat{\phi}_j,\hat{H}]&=&(\sigma h\hat{\bm\phi})_j, \label{eqappb7}\\
\langle[\hat{\cal O}_n,[\hat{\cal O}_n,\hat{\phi}_j]]\rangle&=&0, \label{eqappb8}\\
\langle\{\hat{\cal O}_n,\hat{\phi}_j\}-2\langle\hat{\cal O}_n\rangle\hat{\phi}_j\rangle&=&2\left(\Gamma_\phi O^{\rm T}\right)_{jn}, \label{eqappb9}
\end{eqnarray}
Eq.~(\ref{eqappb6}) allows us to derive the time-evolution equation of the mean value as follows:
\begin{equation}
d{\bm\phi}=\sigma h{\bm\phi}dt+2\Gamma_\phi O^{\rm T}d{\bm W},
\label{eqappb10}
\end{equation}
where $d{\bm W}=\left(dW_1,\dots,dW_N\right)^{\rm T}$. Next, to obtain the equation of motion for the second moment of the Gaussian state, we note that the derivative of the covariance matrix reads
\begin{eqnarray}
&&d\left(\Gamma_\phi\right)_{jk}=\frac{1}{2}\left(\left(d\left\langle\psi_{\rm G}\right|\right)\{\delta\hat{\phi}_j,\delta\hat{\phi}_k\}\left|\psi_{\rm G}\right\rangle\right. \nonumber\\
&&\left.+\left\langle\psi_{\rm G}\right|\{\delta\hat{\phi}_j,\delta\hat{\phi}_k\}\left(d\left|\psi_{\rm G}\right\rangle\right)\right)-d\langle\hat{\phi}_j\rangle d\langle\hat{\phi}_k\rangle.
\label{eqappb11}
\end{eqnarray}
where the last term is the correction term from the It\={o} formula. The first and second terms on the right-hand side of Eq.~(\ref{eqappb11}) are represented by
\begin{eqnarray}
&&\frac{1}{2}\left(\left(d\left\langle\psi_{\rm G}\right|\right)\{\delta\hat{\phi}_j,\delta\hat{\phi}_k\}\left|\psi_{\rm G}\right\rangle+\left\langle\psi_{\rm G}\right|\{\delta\hat{\phi}_j,\delta\hat{\phi}_k\}\left(d\left|\psi_{\rm G}\right\rangle\right)\right) \nonumber\\
&&=\frac{i}{2}\langle[\hat{H},\delta\hat{\phi}_j\delta\hat{\phi}_k+\delta\hat{\phi}_k\delta\hat{\phi}_j]\rangle dt \nonumber\\
&&-\frac{1}{4}\sum_n\langle[\hat{\cal O}_n,[\hat{\cal O}_n,\delta\hat{\phi}_j\delta\hat{\phi}_k+\delta\hat{\phi}_k\delta\hat{\phi}_j]]\rangle dt \nonumber\\
&&+\frac{1}{2}\sum_n\langle\{\hat{\cal O}_n,\delta\hat{\phi}_j\delta\hat{\phi}_k+\delta\hat{\phi}_k\delta\hat{\phi}_j\} \nonumber\\
&&-2\langle\hat{\cal O}_n\rangle(\delta\hat{\phi}_j\delta\hat{\phi}_k+\delta\hat{\phi}_k\delta\hat{\phi}_j)\rangle dW_n.
\label{eqappb12}
\end{eqnarray}
All the terms on the right-hand side of Eq.~(\ref{eqappb12}) are calculated as
\begin{eqnarray}
&&\frac{i}{2}\langle[\hat{H},\delta\hat{\phi}_j\delta\hat{\phi}_k+\delta\hat{\phi}_k\delta\hat{\phi}_j]\rangle=\left(\sigma h\Gamma_\phi\right)_{jk}+(\Gamma_\phi\left(\sigma h\right)^{\rm T})_{jk}, \nonumber\\
\label{eqappb13}\\
&&-\frac{1}{4}\sum_n\langle[\hat{\cal O}_n,[\hat{\cal O}_n,\delta\hat{\phi}_j\delta\hat{\phi}_k+\delta\hat{\phi}_k\delta\hat{\phi}_j]]\rangle=\left(\sigma O^{\rm T}O\sigma^{\rm T}\right)_{jk}, \nonumber\\
\label{eqappb14}\\
&&\langle\{\hat{\cal O}_n,\delta\hat{\phi}_j\delta\hat{\phi}_k+\delta\hat{\phi}_k\delta\hat{\phi}_j\}-2\langle\hat{\cal O}_n\rangle(\delta\hat{\phi}_j\delta\hat{\phi}_k+\delta\hat{\phi}_k\delta\hat{\phi}_j)\rangle \nonumber\\
&&=0, \label{eqappb15}
\end{eqnarray}
where we use Eq.~(\ref{eqappa11}) to derive Eq.~(\ref{eqappb15}), and the last term on the right-hand side of Eq.~(\ref{eqappb11}) is calculated as
\begin{equation}
d\langle\hat{\phi}_j\rangle d\langle\hat{\phi}_k\rangle=4\left(\Gamma_\phi O^{\rm T}O\Gamma_\phi\right)_{jk}.
\label{eqappb16}
\end{equation}
Therefore, we obtain the time-evolution equation of the covariance matrix as follows:
\begin{equation}
\frac{d\Gamma_\phi}{dt}=\sigma h\Gamma_\phi+\Gamma_\phi\left(\sigma h\right)^{\rm T}+\sigma O^{\rm T}O\sigma^{\rm T}-4\Gamma_\phi O^{\rm T}O\Gamma_\phi.
\label{eqappb17}
\end{equation}
We here remark that Eq.~(\ref{eqappb17}) is a deterministic equation in that it does not include stochastic terms, although the Gaussian state obeys the stochastic time evolution described by Eq.~(\ref{eqappb1}). Physically, this means that the envelope of the Gaussian state behaves in the same manner during the evolution in any trajectory provided that the initial state is the same, while the center of the wave packet can fluctuate and exhibit the random motion influenced by continuous monitoring (cf. Eq.~(\ref{eqappb10})). In particular, we note that Eq.~(\ref{eqappb17}) has a unique steady solution $d\Gamma_\phi/dt=0$ in the limit of $t\rightarrow\infty$, independent of initial states.

\begin{figure}[]
\includegraphics[width=8.5cm]{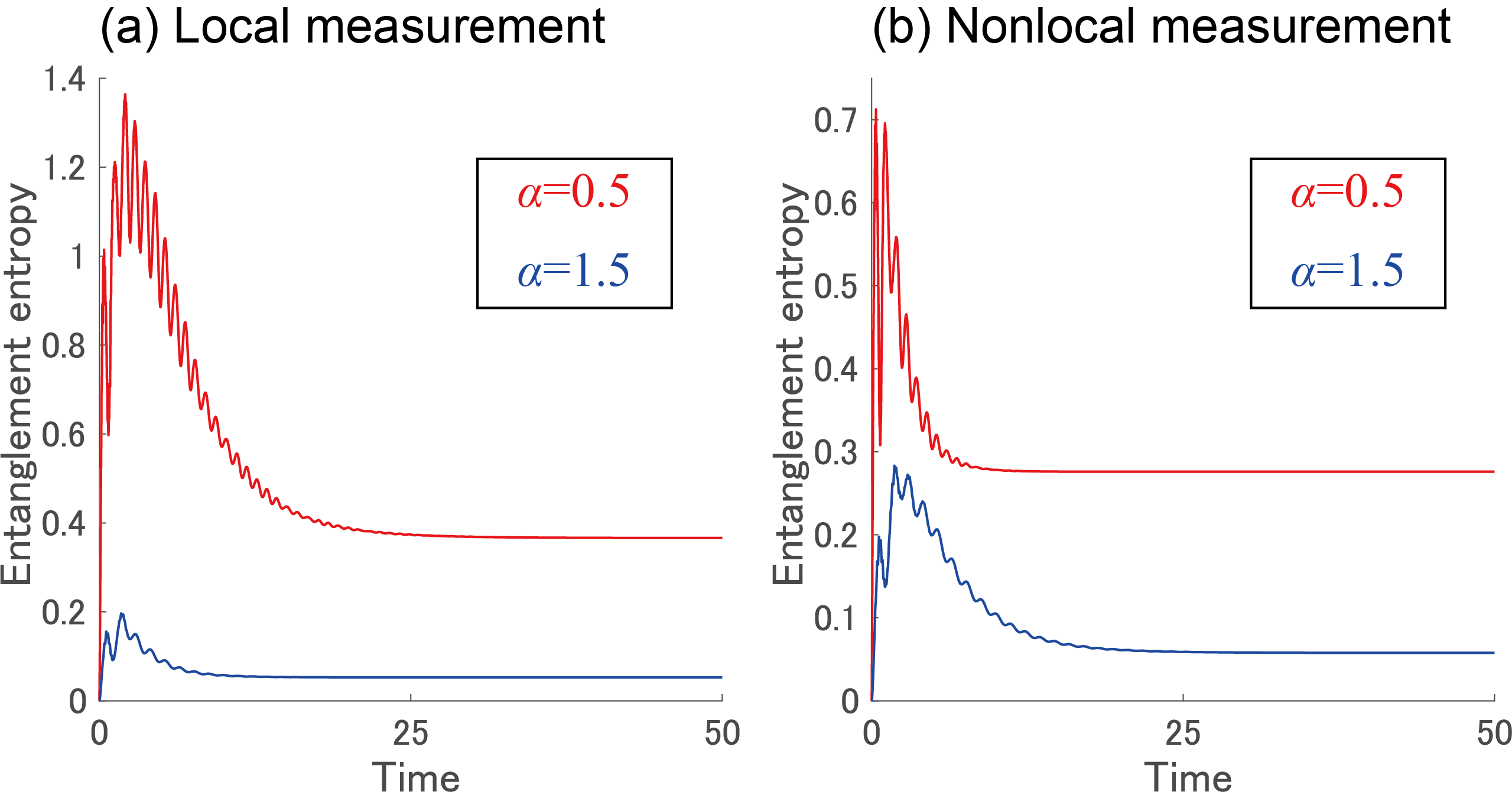}
\caption{\label{figappb}Time evolution of the half-chain entanglement entropy under the (a) local or (b) nonlocal measurement. We set the system size to be $L=100$ in both cases and the measurement strength as $\gamma=0.5$ in (a) and $\gamma=0.03$ in (b).}
\end{figure}

As examples, we show the time evolution of the entanglement entropy under the local or nonlocal measurement by simulating Eq.~(\ref{eqappb17}) and calculating the symplectic eigenvalue of the covariance matrix [Fig.~\ref{figappb}]. We note that $\Gamma_\phi\left(t=0\right)=\left(1/2\right)1_{2N}$ is used as an initial state. We confirm that, after the initial transient evolution, the entanglement entropy converges to a constant value.

%
%

\section{\label{secC}Non-Hermitian dynamics}
\begin{figure}[]
\includegraphics[width=8.5cm]{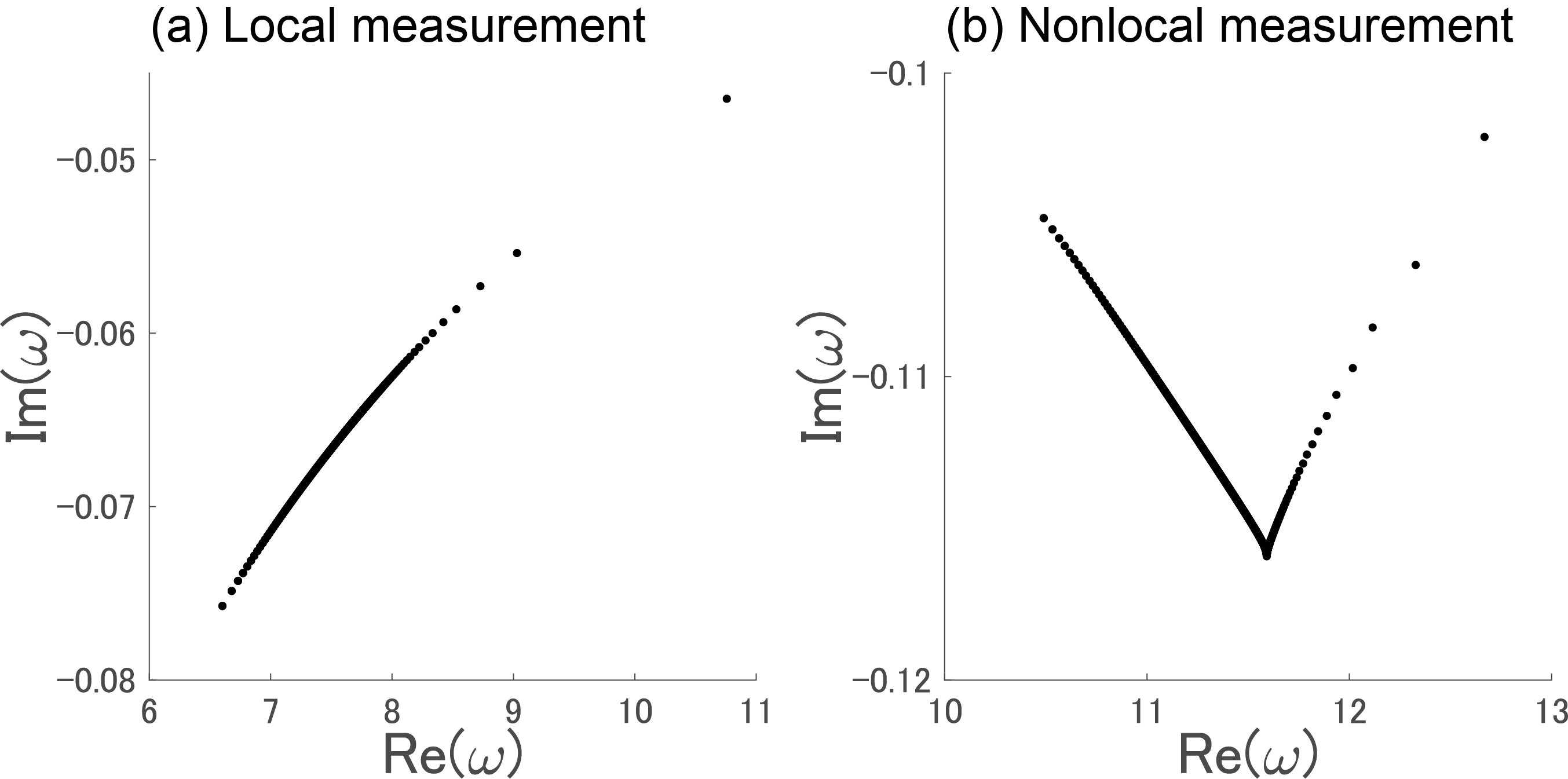}
\caption{\label{figappc}Excitation spectrum of the effective non-Hermitian Hamiltonian under the (a) local or (b) nonlocal measurement. We set the system size to be $L=500$ in both cases and $\alpha=0.5$ and $\gamma=0.5$ in (a) and $\alpha=0.3$ and $\gamma=0.01$ in (b).}
\end{figure}

In this section, we derive an effective non-Hermitian Hamiltonian that exactly describes the time evolution for the covariance matrix of a bosonic Gaussian state under measurements (see also Refs.~\cite{Minoguchi2022,Young2024} for related discussions). Specifically, we introduce the non-Hermitian Hamiltonian,
\begin{equation}
\hat{H}_{\rm eff}=\hat{H}-\frac{i}{2}\hat{\Delta},
\label{eqappc1}
\end{equation}
where $\hat{H}$ is given in Eq.~(\ref{eqappb2}), and
\begin{equation}
\hat{\Delta}=\sum_{j,k}D_{jk}\hat{\phi}_j\hat{\phi}_k.
\label{eqappc2}
\end{equation}
The non-Hermitian Hamiltonian leads to the (nonlinear) time evolution of the density operator written as
\begin{equation}
\hat{\rho}\left(t\right)=\frac{e^{-i\hat{H}_{\rm eff}t}\hat{\rho}\left(0\right)e^{i\hat{H}_{\rm eff}^\dag t}}{{\rm Tr}(e^{i\hat{H}_{\rm eff}^\dag t}e^{-i\hat{H}_{\rm eff}t}\hat{\rho}\left(0\right))},
\label{eqappc3}
\end{equation}
and its time derivative gives
\begin{equation}
\frac{d\hat{\rho}}{dt}=-i[\hat{H},\hat{\rho}]-\frac{1}{2}\{\hat{\Delta},\hat{\rho}\}+\hat{\rho}{\rm Tr}(\hat{\Delta}\hat{\rho}).
\label{eqappc4}
\end{equation}
The corresponding equation of motion for the covariance matrix then reads
\begin{equation}
\frac{d\Gamma_\phi}{dt}=\sigma h\Gamma_\phi+\Gamma_\phi\left(\sigma h\right)^{\rm T}+\frac{1}{2}\sigma D\sigma^{\rm T}-2\Gamma_\phi D\Gamma_\phi.
\label{eqappc5}
\end{equation}
When identifying $D=2O^{\rm T}O$, the above equation precisely matches the time-evolution equation of the covariance matrix of a measured quantum state, as derived in Eq.~(\ref{eqappb17}). Thus, we can conclude that the dynamics of the covariance matrix at each quantum trajectory is equivalent to the time evolution governed by the following non-Hermitian Hamiltonian,
\begin{equation}
\hat{H}_{\rm eff}=\hat{H}-i\sum_n\hat{\cal O}_n^2.
\label{eqappc6}
\end{equation}
Since the measurement operators are given in linear combinations of the position operators in our setup, the non-Hermitian Hamiltonian can be written in the following form:
\begin{equation}
\hat{H}_{\rm eff}=\sum_j\frac{\Omega}{2}\hat{p}_j^2+\sum_{j,k}\frac{A_{jk}}{2}\hat{x}_j\hat{x}_k,
\label{eqappc7}
\end{equation}
where $A$ is a complex-valued symmetric matrix. We note that the spectrum of the non-Hermitian Hamiltonian is calculated from the eigenvalue of the matrix $A$. As examples, we show the non-Hermitian spectrum under the local or nonlocal measurement in Fig.~\ref{figappc}.

We comment on a steady state under a Gaussian measurement, where a system Hamiltonian has a quadratic form $\hat{\bm\phi}$, and measurement operators are written in linear combinations of $\hat{\bm\phi}$. In this case, the stationary state is the squeezed vacuum state regardless of a choice of initial states because all the imaginary parts of the spectrum of the non-Hermitian Hamiltonian are negative as shown in Fig.~\ref{figappc}, ensuring that the components of any other states should exponentially decay in time. Thus we conclude that an arbitrary initial state converges to a steady Gaussian state under a Gaussian measurement.

%
%

\section{\label{secD}Spatial correlation}
\begin{figure}[]
\includegraphics[width=8.5cm]{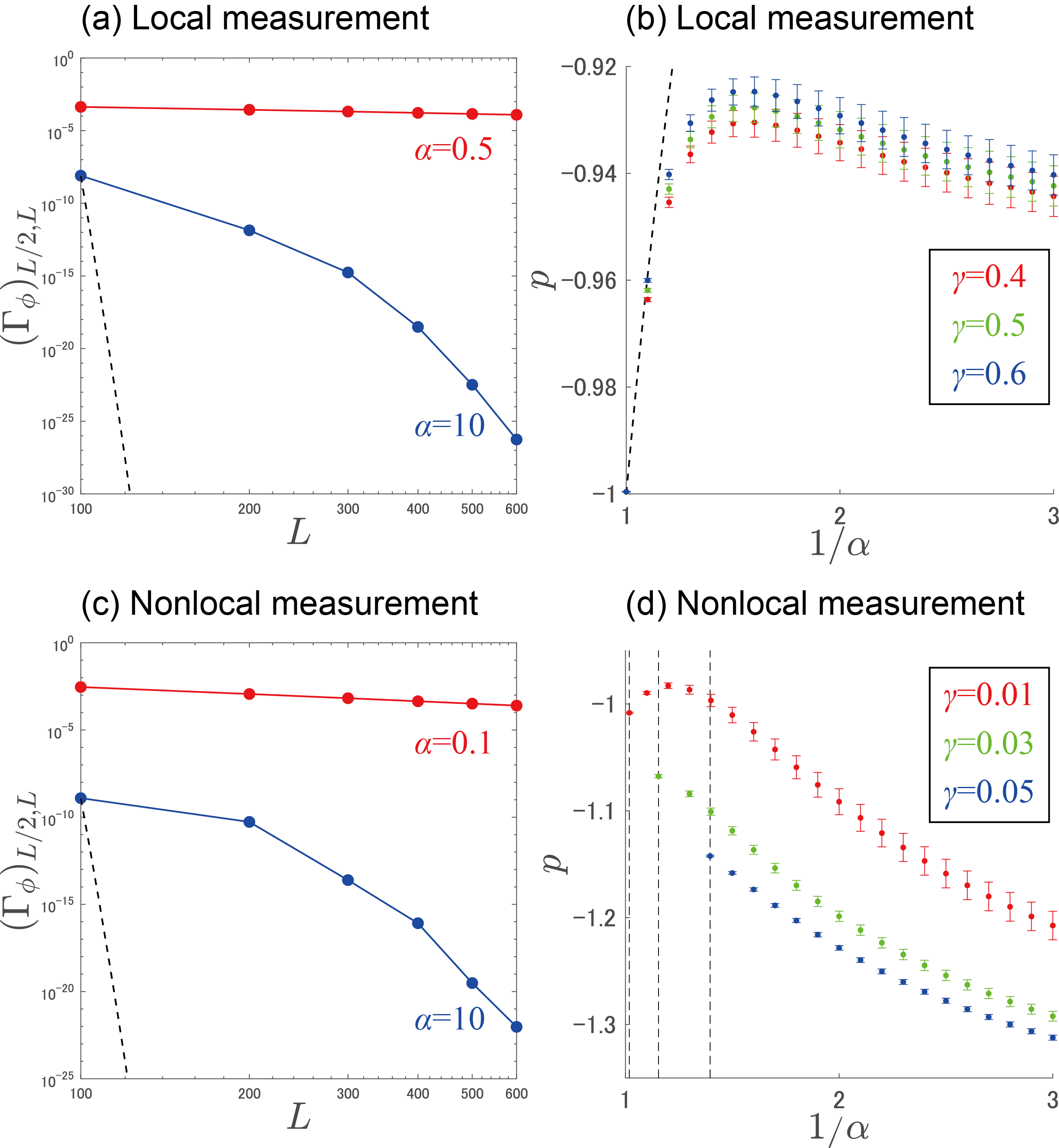}
\caption{\label{figappd}Scaling of the correlation function under the [(a) and (b)] local or [(c) and (d)] nonlocal measurement. (a) Spatial correlation in the subvolume-law phase (red, $\alpha=0.5$ and $\gamma=10$) and the area-law phase (blue, $\alpha=10$ and $\gamma=10$) in a semilog plot. (b) Size-scaling exponent $p$ in the subvolume-law phase. We show the black dashed line proportional to $1/\alpha$. (c) Spatial correlation in the subvolume-law phase (red, $\alpha=0.1$ and $\gamma=0.01$) and the area-law phase (blue, $\alpha=10$ and $\gamma=1.5$) in a semilog plot. (d) Size-scaling exponent $p$ in the subvolume-law phase. We show the black dashed lines as the transition points. In (a) and (b), the black dashed lines are obtained by Eq.~(\ref{eqappd8}).}
\end{figure}
In this section, we discuss the spatial correlation in the subvolume- and area-law phases. The correlation between $\hat{x}_j$ and $\hat{x}_k$ is given by the $\left(j,k\right)$ element of the steady solution of Eq.~(\ref{eqappb17}). Furthermore, the correlation function obeys $\left(\Gamma_\phi\right)_{jk}=f\left(\left|j-k\right|\right)$ due to the spatial periodicity (i.e., translational invariance) of the system. Thus we show the component of the correlation function between $j=L/2$ (the right end of the subregion $A$) and $k=L$ (the right end of the residual region) in Figs.~\ref{figappd}(a) and \ref{figappd}(c). We confirm that, in the subvolume-law phase, the correlation functions exhibit the power-law decay, explicitly written as
\begin{equation}
\left(\Gamma_\phi\right)_{L/2,L}\propto L^{-p},
\label{eqappd1}
\end{equation}
while they exhibit the spatially exponential decay in the area-law phase as we discuss in the main text.

We show the dependence of the exponent $p$ on the parameters $\alpha$ and $\gamma$ in Figs.~\ref{figappd}(b) and \ref{figappd}(d). Under the local measurement, $p$ as a function of $\alpha$ exhibits the nonmonotonic dependence mainly independent of $\gamma$. Meanwhile, $p$ is sensitive to the changes in both $\alpha$ and $\gamma$ under the nonlocal measurement. In contrast, a previous study on free-fermion systems with long-range couplings has proposed that the scaling of the correlation function obeys $L^{-a-1/2}$, where the exponent $a$ characterizes the power-law decay of the long-range coupling in the Hamiltonian and is independent of the measurement strength~\cite{Muller2022}. These results indicate that the critical behavior found in the subvolume-law phase of the present setup is different from the one proposed by the previous work; this point further suggests that our findings go beyond the Tomonaga-Luttinger liquid-type descriptions.

To provide analytical insights into the spatially exponential decay of the correlation function, we can perform an analytical analysis of the correlation-function scaling in the area-law phase. We note that the non-Hermitian spectrum in the case of the local measurement is given by
\begin{equation}
\omega\left(q\right)=\sqrt{\Omega\left(\Omega+Kh_\alpha\left(q\right)-2i\gamma\right)},~\left(q\in{\mathbb R}\right),
\label{eqappd2}
\end{equation}
and the one in the case of the nonlocal measurement is given by
\begin{equation}
\omega\left(q\right)=\sqrt{\Omega\left(\Omega+Kh_\alpha\left(q\right)-8i\gamma{\rm Li}_\alpha\left(1\right)\right)},~\left(q\in{\mathbb R}\right),
\label{eqappd3}
\end{equation}
where
\begin{equation}
h_\alpha\left(q\right)=2{\rm Li}_\alpha\left(q\right)-{\rm Li}_\alpha\left(e^{iq}\right)-{\rm Li}_\alpha\left(e^{-iq}\right),
\label{eqappd4}
\end{equation}
and ${\rm Li}_\alpha\left(z\right)$ is a polylogarithm function defined as
\begin{equation}
{\rm Li}_\alpha\left(z\right)=\sum_{r=1}^\infty\frac{z^r}{r^\alpha}.
\label{eqappd5}
\end{equation}
We note that ${\rm Li}_\alpha\left(1\right)$ is a zeta function, and it has a pole at $\alpha=1$. The key observation is that, when $\alpha>1$, the correlation function can be obtained by
\begin{equation}
\left(\Gamma_\phi\right)_{jk}\propto\int dq\frac{e^{iq\left|j-k\right|}}{{\rm Re}\left(\omega\left(q\right)\right)}.
\label{eqappd6}
\end{equation}
This formula allows us to derive the asymptotic analytical form of the correlation function for $\left|j-k\right|\gg1$. In the case of the local measurement, by using the approximation,
\begin{equation}
h_\alpha\left(q\right)\simeq4\sin^2q\simeq4q^2,
\label{eqappd7}
\end{equation}
when $\alpha\gg1$, we can rewrite Eq.~(\ref{eqappd4}) as
\begin{eqnarray}
\left(\Gamma_\phi\right)_{jk}&\simeq&\int dq\frac{e^{iq\left|j-k\right|}}{\sqrt{1+\kappa^2q^2}} \nonumber\\
&\simeq&K_0\left(\left|j-k\right|/\kappa\right) \nonumber\\
&\simeq&\frac{e^{-\left|j-k\right|/\kappa}}{\sqrt{\left|j-k\right|/\kappa}},
\label{eqappd8}
\end{eqnarray}
where $K_0\left(x\right)$ is the modified Bessel function of the second kind of $0$th order, and $\kappa$ is defined as
\begin{equation}
\kappa=\frac{K}{\sqrt{\Omega^2+4\gamma^2}}.
\label{eqappd9}
\end{equation}
Meanwhile, in the case of the nonlocal measurement, the similar asymptotic form of the correlation function can also be obtained by the replacement $\gamma\rightarrow4\gamma$. This analysis qualitatively explains the exponential decay of the correlation function in the area-law phase. However, we also find that there is still quantitative discrepancy between the analytical results and numerical results as shown in Figs.~\ref{figappd}(a) and \ref{figappd}(c), whose comprehensive understanding remains as a future problem.

We comment on the possibility of providing the analytical asymptotic form of the correlation function when $\alpha\gtrsim1$. For this purpose, we can use the Taylor expansion of $h_\alpha\left(q\right)$ in the derivation of Eq.~(\ref{eqappd7}). However, when $\alpha\gtrsim1$, this approximation breaks down around $q=0$. Thus the singularity makes it difficult to analytically and numerically calculate the asymptotic form.

%
%

\section{\label{secE}Entanglement growth rate and phase transition}
\begin{figure}[]
\includegraphics[width=8.5cm]{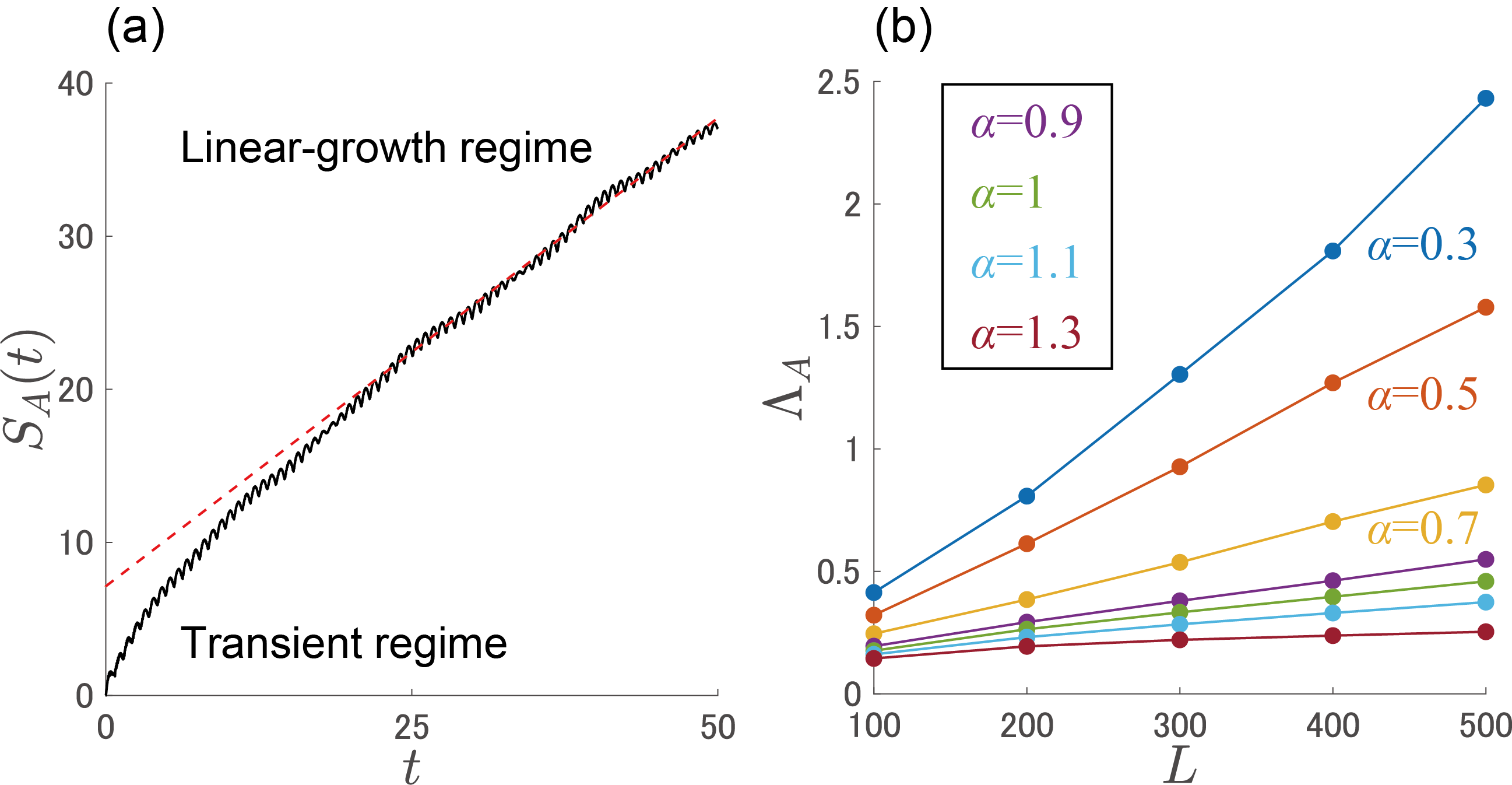}
\caption{\label{figappe}Entanglement growth under the long-range coupling. (a) Time evolution of the half-chain entanglement entropy. The red dashed line represents a linear fit of the entanglement entropy. The parameters are set to be $\alpha=0.5$ and $L=200$. (b) System-size scaling of the growth rate $\Lambda_{A}$.}
\end{figure}
In this section, we first discuss the growth rate of the entanglement entropy under unitary time evolution, following the discussions in Refs.~\cite{Hackl2018,Bianchi2018}. We find that the half-chain entanglement entropy in free-boson systems linearly increases in time with a rate $\Lambda_A$ after an initial transient regime [Fig.~\ref{figappe}(a)]. We can extract the growth rate in our setup from a linear fit, and the obtained results are shown in Fig.~\ref{figappe}(b). The key point is that the growth rate is proportional to the system size when $0<\alpha<1$, while it becomes constant for various system sizes when $\alpha>1$. The finding means that the growth rate diverges in the thermodynamic limit when $0<\alpha<1$. 

Remarkably, we can provide an analytical argument to explain why the boundary between the subvolume- and area-law phases lies in the $\alpha=1$ line under the local measurement in terms of the entanglement growth rate above. The key observation is that, in the thermodynamic limit, the growth rate of the entanglement entropy diverges when $0<\alpha<1$, and the local measurement cannot suppress the entanglement generation at any measurement strength. To understand the origin of the divergence of the growth rate, we express the spectrum of the system Hamiltonian from the form of Eq.~(\ref{eqappc7}) as
\begin{equation}
\omega\left(q\right)=\sqrt{\Omega\left(\Omega+Kh_\alpha\left(q\right)\right)},~\left(q\in{\mathbb R}\right),
\label{eqappe1}
\end{equation}
where $h_\alpha\left(q\right)$ is given in Eq.~(\ref{eqappd4}). The growth rate $\Lambda_A$ is calculated from the summation of the Lyapunov exponents of all the eigenvalues~\cite{Bianchi2018}. Here, the Lyapunov exponent is given by $\lambda\left(q\right)={\rm Im}\left(\omega\left(q\right)\right)$. Remarkably, when $0<\alpha<1$, $\omega\left(q\right)$ becomes purely imaginary because $h_\alpha\left(q\right)$ always takes negative values. Therefore, the infinite number of the eigenmodes contribute to the growth rate of the entanglement entropy, which leads to the divergent rate of the entanglement generation. The physical insight is that, in the thermodynamics limit, all the eigenmodes become unstable due to the long-range coupling~\footnote{Although a Hamiltonian satisfies the Hermiticity $\hat{H}^\dag=\hat{H}$, it can effectively have complex-valued eigenvalues when nonsquare-integrable eigenfunctions are included (see, e.g., Sec.~4.2 in Ref.~\cite{Ashida2020}) Physically, these states correspond to unstable resonances that have finite lifetimes; the simplest example for this is a quantum particle on an inverted harmonic potential.}.

To include the effect of measurements, we perform an analytical analysis of the non-Hermitian dynamics. In the case of the local measurement, one can calculate the non-Hermitian spectrum given in Eq.~(\ref{eqappd2}). We remark that the term proportional to $i\gamma$ in Eq.~(\ref{eqappd2}) is independent of $\alpha$ and does not cause singularity, but simply shifts all the eigenvalues by the constant value in the imaginary-axis direction. This indicates that the transition cannot be induced by increasing the measurement strength, which is consistent with our numerical observation. Meanwhile, Eq.~(\ref{eqappd2}) still possesses the same spectral singularity at $\alpha=1$ as discussed in the Hermitian case above. Thereby, the subvolume- and area-law phases are separated at $\alpha=1$ at any measurement strength.

Finally, we mention that the phase transition in the case of the nonlocal measurement might be also understood from a similar analytical argument based on the non-Hermitian dynamics, while we leave its comprehensive understanding for future work.

%
%

\section{\label{secF}Short-range coupling and nonlocal measurement}
\begin{figure}[]
\includegraphics[width=8.5cm]{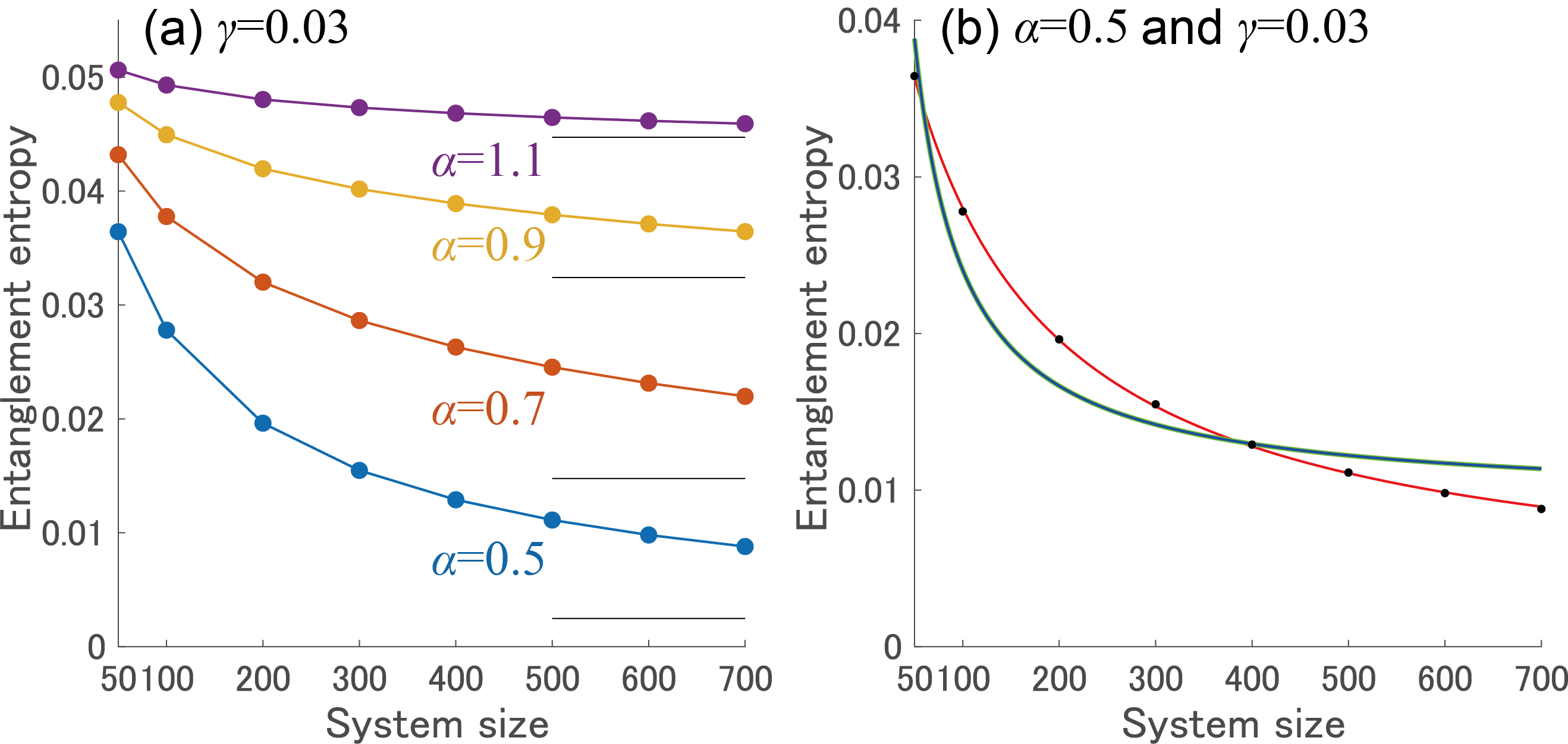}
\caption{\label{figappf}Half-chain entanglement entropy in the setup where the coupling is short-ranged and the system is subject to the nonlocal measurement. (a) System-size dependence of the entanglement entropy. We show the asymptotic values of the entanglement entropy by the black horizontal lines extracted from the Pad\'{e} approximant with $n=1$. (b) Pad\'{e} approximant with $n=1$ (red), $n=2$ (green), and $n=3$ (blue).}
\end{figure}
In this section, we consider the setup with the short-range particle-particle coupling under the nonlocal measurement and show the scaling of the entanglement entropy. We focus on the entanglement entropy of the bosonic Gaussian state in the long-time limit similar to the above case. The considered setup obeys Eq.~(\ref{eqappb1}), where the Hamiltonian is
\begin{equation}
\hat{H}=\sum_{j=1}^L\frac{\Omega}{2}(\hat{p}_j^2+\hat{x}_j^2)+\sum_{j=1}^{L-1}\frac{K}{2}(\hat{x}_j-\hat{x}_{j+1})^2,
\label{eqappf1}
\end{equation}
and the measurement operators are given by
\begin{equation}
\hat{\cal O}_n=\sqrt{\frac{\gamma}{r^\alpha}}(\hat{x}_j\pm\hat{x}_{j+r})
\label{eqappf2}
\end{equation}
with $n=\left(j,r,\pm\right)$. For the sake of simplicity, we set $K=\Omega=1$. Figure~\ref{figappf}(a) shows the resulting entanglement scaling at different $\alpha$. To extract the asymptotic behavior of the entanglement entropy in the thermodynamic limit, we apply the Pad\'{e} approximant, which is the best approximation of a function by a rational function~\cite{Baker1996}. Specifically, the Pad\'{e} approximant of $n$th order is given by
\begin{equation}
R\left(x\right)=\frac{\displaystyle\sum_{k=0}^na_kx^k}{1+\displaystyle\sum_{k=1}^nb_kx^k}.
\label{eqappf3}
\end{equation}
We show the Pad\'{e} approximant up to third order in Fig.~\ref{figappf}(b). One can infer that Pad\'{e} approximant with first order is the best fit. We then find that, at any $\alpha$, the entanglement entropy asymptotically converges a constant in the limit of $L\rightarrow\infty$ as indicated by black horizontal lines [Fig.~\ref{figappf}(a)]. Thus, we conclude that the entanglement entropy in the setup always obeys the area law.

%
%

\section{\label{secG}Finite-size scaling analysis}
\begin{figure}[]
\includegraphics[width=8.5cm]{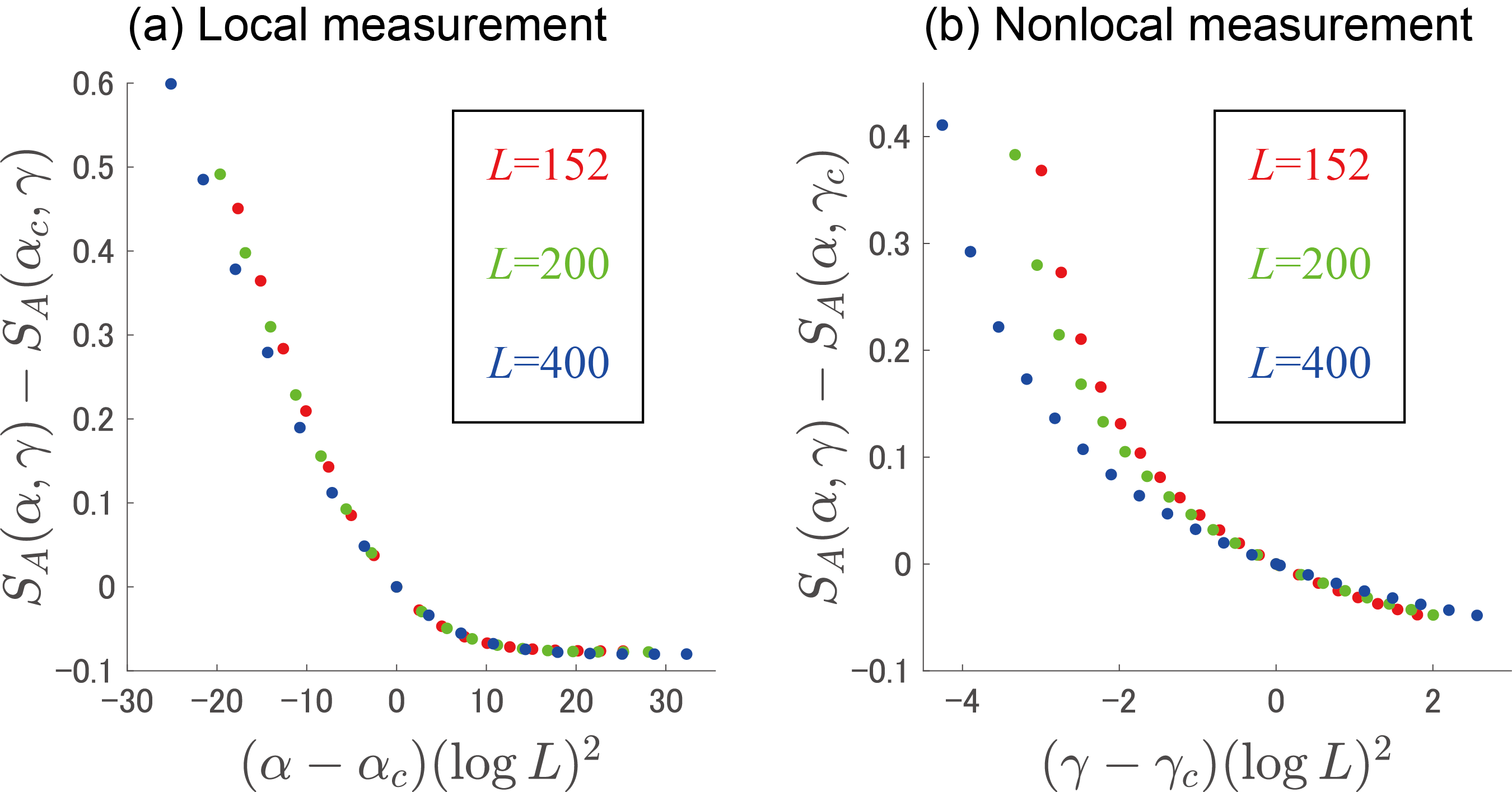}
\caption{\label{figappg1}Data set of the half-chain entanglement entropy under the (a) local or (b) nonlocal measurement rescaled in the Berezinskii-Kosterlitz-Thouless scaling in Eqs.~(\ref{eqappg1}) and (\ref{eqappg2}). We set $\gamma=0.5$ in (a) or $\alpha=0.3$ in (b).}
\end{figure}
\begin{figure}[]
\includegraphics[width=8.5cm]{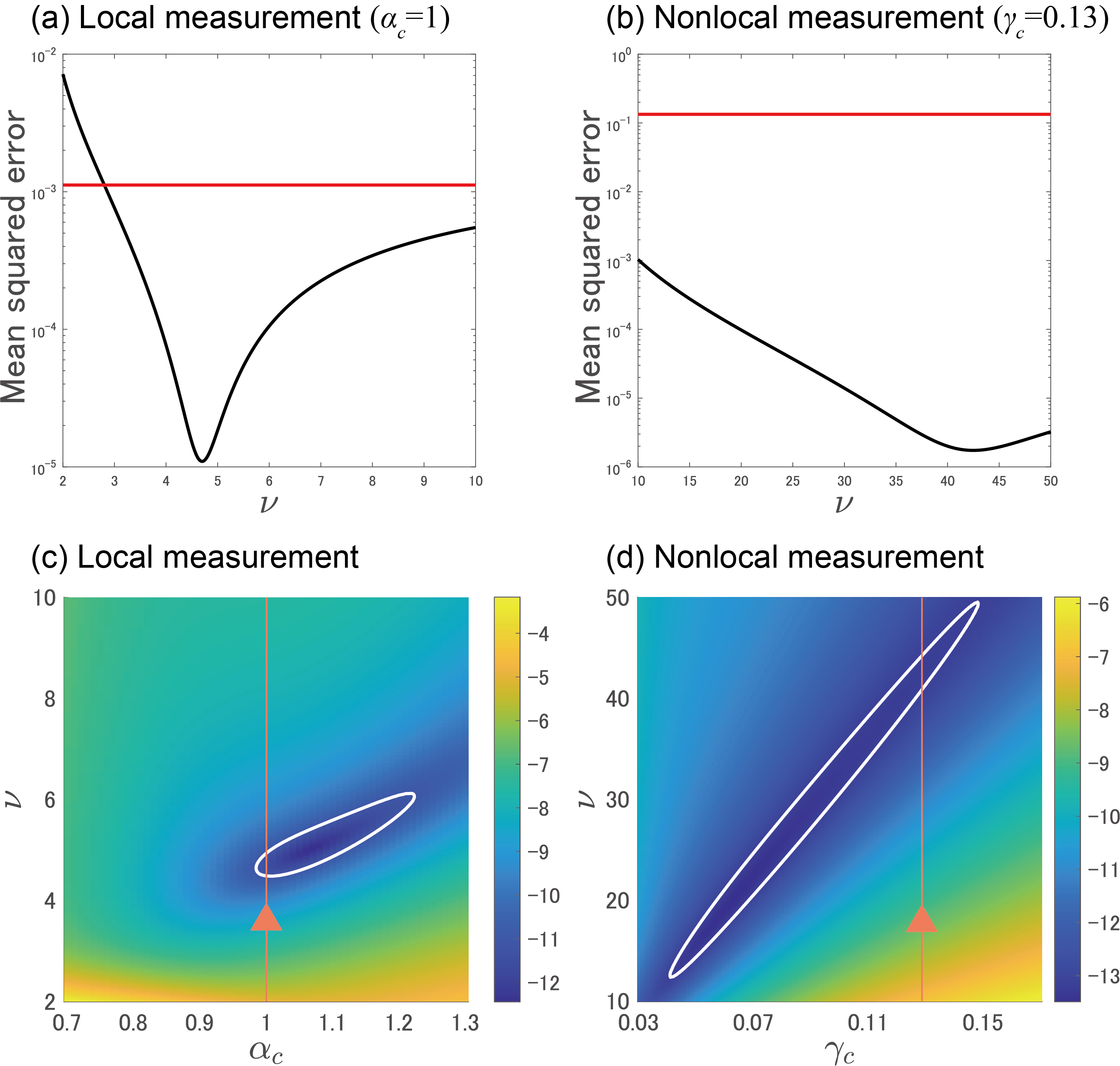}
\caption{\label{figappg2}Mean-squared error under the [(a) and (c)] local or [(b) and (d)] nonlocal measurement. [(a) and (b)] Mean squared error in the proposed scenario (black) and in the Berezinskii-Kosterlitz-Thouless scenario (red) in a semilog plot. [(c) and (d)] Color map of the global mean squared error in a logarithmic scale. The white loops in (c) and (d) represent the contour lines with $4\min{\cal E}$ and $1.4\min{\cal E}$, respectively. The orange lines match the critical points. We set $\gamma=0.5$ in [(a) and (c)] or $\alpha=0.3$ in [(b) and (d)].}
\end{figure}
In this section, we describe the details about how we analyze the universality class of the entanglement transition under the local or nonlocal measurement. We compare the rescaled date sets of the half-chain entanglement entropy to assess whether the scenario proposed in our work or the Berezinskii-Kosterlitz-Thouless (BKT) scenario, which has been shown to be valid for several measurement-induced phase transitions in free fermions~\cite{Alberton2021,Minato2022,Muller2022}, would be more feasible one. In the proposed scenario, the entanglement entropy in the vicinity of the transition point is assumed to obey Eq.~(\ref{eq6}) in the case of the local measurement or Eq.~(\ref{eq7}) in the case of the nonlocal measurement. We note that the similar scaling function has been proposed by Ref.~\cite{Fuji2020}. Meanwhile, in the BKT scenario, the entanglement entropy obeys~\cite{Harada1997,Carrasquilla2012,Alberton2021}
\begin{equation}
S_A\left(\alpha,\gamma\right)-S_A\left(\alpha_c,\gamma\right)=G\left[\left(\alpha-\alpha_c\right)\left(\ln L\right)^2\right]
\label{eqappg1}
\end{equation}
in the case of the local measurement or
\begin{equation}
S_A\left(\alpha,\gamma\right)-S_A\left(\alpha,\gamma_c\right)=G\left[\left(\gamma-\gamma_c\right)\left(\ln L\right)^2\right]
\label{eqappg2}
\end{equation}
in the case of the nonlocal measurement, where $G\left(x\right)$ is a smooth function. Figures~\ref{figappg1}(a) and \ref{figappg1}(b) show the rescaled data sets of the entanglement entropy assuming the BKT scaling in Eqs.~(\ref{eqappg1}) and (\ref{eqappg2}). We recall that, in both local and nonlocal measurements, the data set of the entanglement entropy collapses into a single curve in the proposed scaling [Figs.~\ref{fig4}(c) and \ref{fig4}(d)]. Meanwhile, Figs.~\ref{figappg1}(a) and \ref{figappg1}(b) clearly show that the data fail to reproduce the BKT scaling.

To qualitatively estimate how well the numerical results fit the proposed scaling, we evaluate the mean squared error defined by
\begin{equation}
{\cal E}=\frac{1}{n}\sum_{i=1}^n\left(y_i-Q\left(x_i\right)\right)^2,
\label{eqappg3}
\end{equation}
where $x_i$ is the $i$th scaling variable, $y_i$ denotes the $i$th value in the data sets in the cases of $L=152$ and $L=400$, $n$ is the total number of the data points. We note that the function $Q\left(x\right)$ is approximately determined from the data set for $L=200$. In the proposed scaling, $Q\left(x\right)$ is assumed to be the polynomial approximation of the scaling function $F\left(x\right)$ given in Eqs.~(\ref{eq6}) and (\ref{eq7}). The black curves in Fig.~\ref{figappg2}(a) and \ref{figappg2}(b) show the mean squared error as a function of $\nu$. Meanwhile, in the BKT scaling, it is the polynomial approximation of the scaling function $G\left(x\right)$ given in Eqs.~(\ref{eqappg1}) and (\ref{eqappg2}). The red horizontal lines in Figs.~\ref{figappg2}(a) and \ref{figappg2}(b) also show the mean squared error in this case. We remark that the entanglement entropy obeys the proposed scaling because the black curve takes a minimum value at a specific value of $\nu$. Furthermore, we calculate the global mean squared error in the proposed scaling to estimate the error of $\nu$ [Figs.~\ref{figappg2}(c) and \ref{figappg2}(d)]. The white loop surrounds the region around the minimum means squared error ${\cal E}_{\rm min}$ satisfying ${\cal E}<4{\cal E}_{\rm min}$ in Fig.~\ref{figappg2}(c) or ${\cal E}<1.4{\cal E}_{\rm min}$ in Fig.~\ref{figappg2}(d). Finally, all the plots allow us to extract $\nu=4.7\pm0.3$ in the case of the local measurement or $\nu=42\pm3$ in the case of the nonlocal measurement. We thus argue that the most probable conclusion from the obtained results is that the measurement-induced phase transition should belong to an unconventional universality class different from the BKT class.

%
%

\section{\label{secH}Imperfect measurement}
\begin{figure}[]
\includegraphics[width=8.5cm]{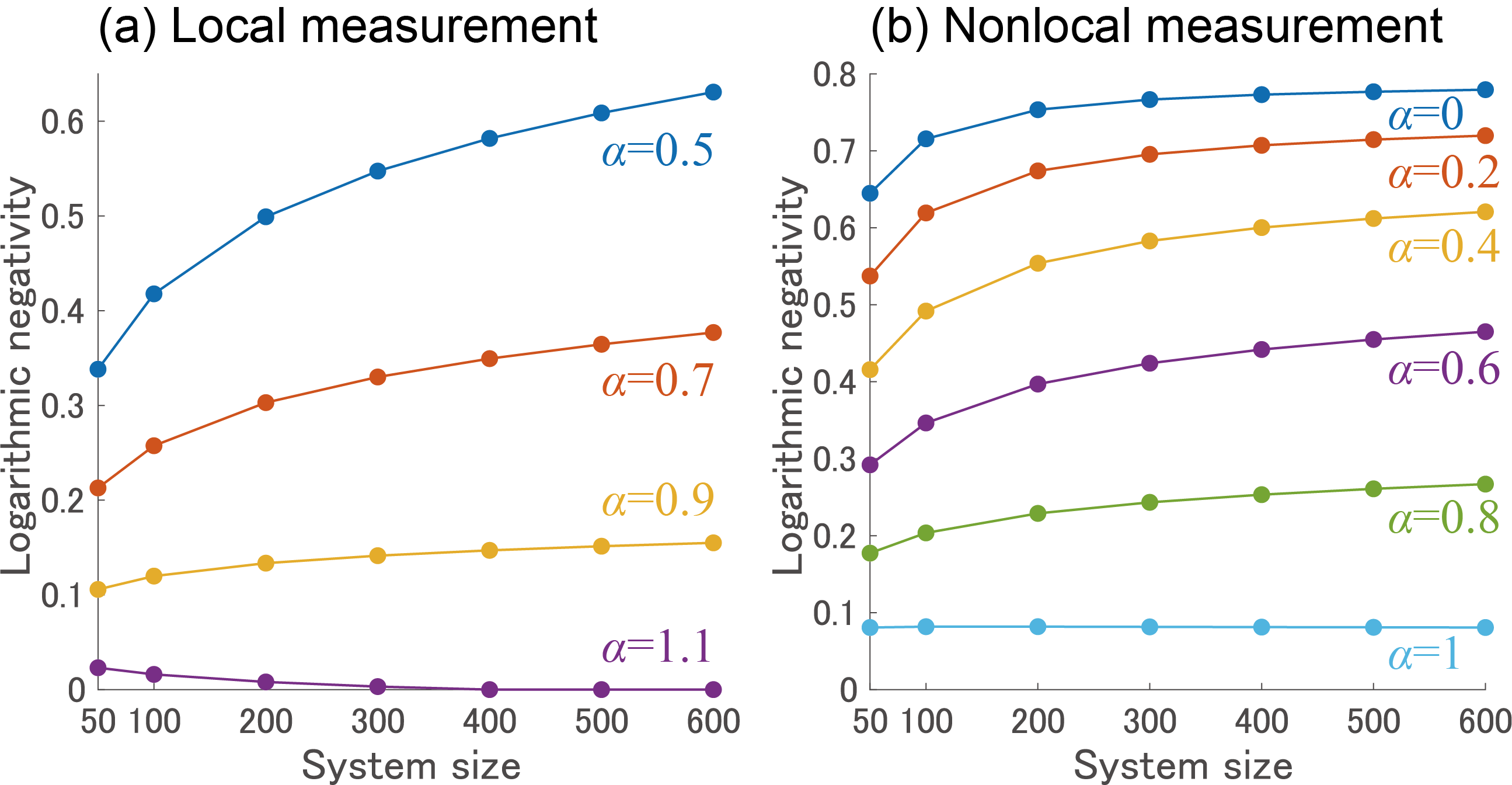}
\caption{\label{figapph1}Half-chain logarithmic negativity under the (a) local or (b) nonlocal measurement. We set the measurement efficiency as $\eta=0.6$ in both cases and the measurement strength as $\gamma=0.5$ in (a) and $\gamma=0.01$ in (b).}
\end{figure}
\begin{figure}[]
\includegraphics[width=5cm]{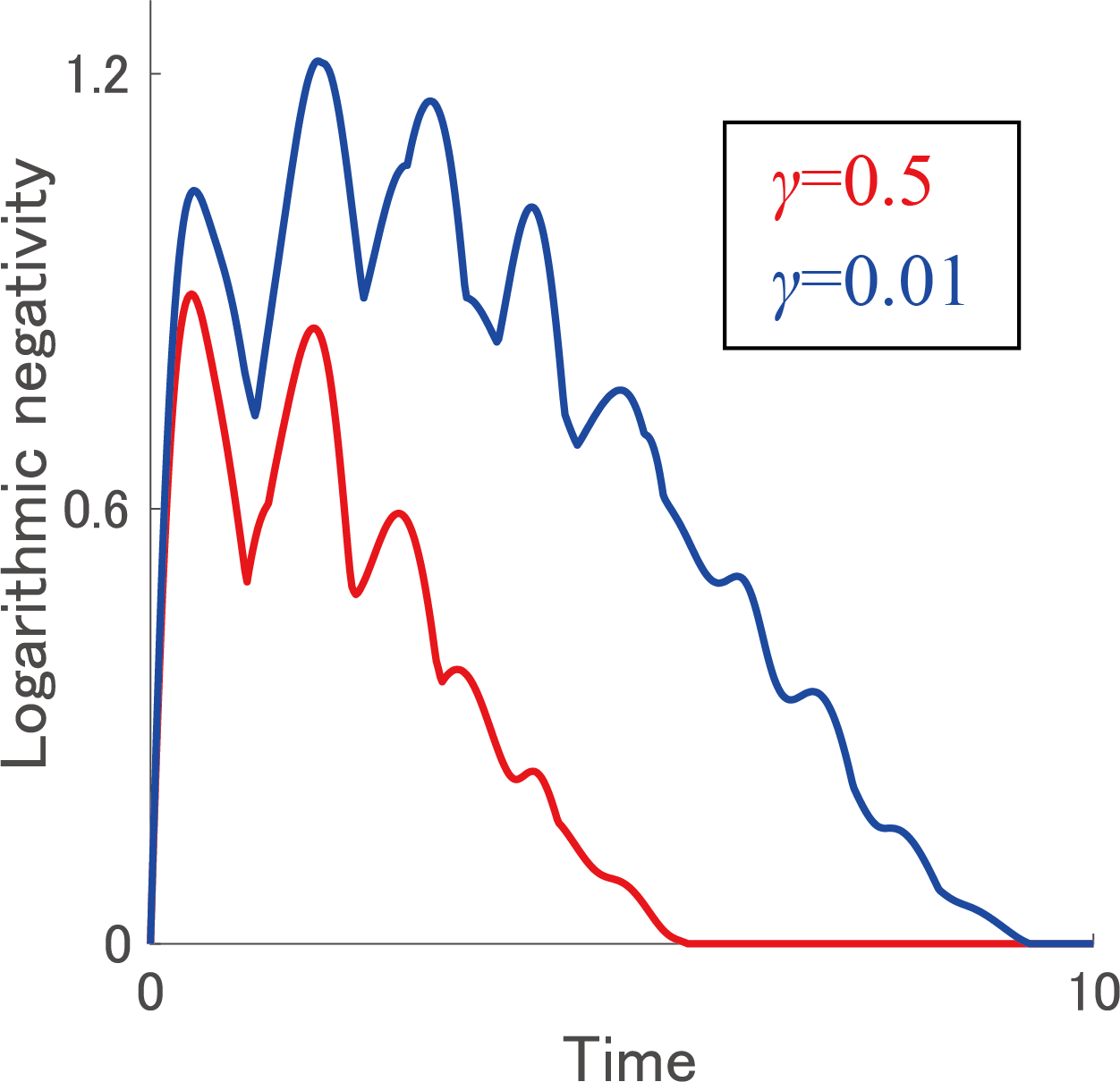}
\caption{\label{figapph2}Time evolutions of the half-chain logarithmic negativity under the local (red) or nonlocal (blue) dissipation. We set $L=100$ and $\alpha=0.7$ in both cases.}
\end{figure}
In this section, we investigate the entanglement behavior of our setup under imperfect measurements, where the steady state evolves into a mixed state. We note that, in this case, entanglement in a mixed state can be characterized by the logarithmic negativity rather than the entanglement entropy for a pure state. Thus our aim is to calculate the half-chain logarithmic negativity in our setup under imperfect measurement, where the time-evolution of the covariance matrix reads~\cite{Minoguchi2022}
\begin{equation}
\frac{d\Gamma_\phi}{dt}=\sigma h\Gamma_\phi+\Gamma_\phi\left(\sigma h\right)^{\rm T}+\sigma O^{\rm T}O\sigma^{\rm T}-4\eta\Gamma_\phi O^{\rm T}O\Gamma_\phi.
\label{eqapph1}
\end{equation}
Here, $0\leq\eta\leq1$ represents the measurement efficiency.

We first review the way to calculate logarithmic negativity of a bosonic (mixed) Gaussian state with $N$ bosonic modes $\left(N\in2{\mathbb N}\right)$. The logarithmic negativity is defined as
\begin{equation}
{\cal N}_A=\ln||\hat{\rho}^{\rm T_{\it \bar{A}}}||_1,
\label{eqapph2}
\end{equation}
where $\hat{\rho}^{\rm T_{\it \bar{A}}}$ represents the partial transpose of the original density matrix with respect to the subregion $A$, and $||\hat{\rho}^{\rm T_{\it \bar{A}}}||_1$ represents the trace norm of $\hat{\rho}^{\rm T_{\it \bar{A}}}$~\cite{Vidal2002}. We remark that, for a bosonic Gaussian state, the logarithmic negativity only depends on the covariance matrix. Indeed, by defining
\begin{equation}
{\cal T}_{\bar{A}}=1_N\oplus\left(1_{N/2}\oplus\left(-1_{N/2}\right)\right),
\label{eqapph3}
\end{equation}
where $1_M$ is an $M\times M$ identity matrix, the partial transpose for the original covariance matrix $\Gamma_\phi$ is performed as
\begin{equation}
\tilde{\Gamma}_\phi={\cal T}_{\bar{A}}\Gamma_\phi{\cal T}_{\bar{A}}.
\label{eqapph4}
\end{equation}
Let $\left(\tilde{\kappa}_1,\dots,\tilde{\kappa}_N\right)$ denote the symplectic eigenvalues of the partially transposed covariance matrix. We then obtain the logarithmic negativity of the Gaussian state as follows~\cite{Adesso2007}:
\begin{equation}
{\cal N}_A=\sum_{l=1}^{N}\ln{\rm max}\left(1,\frac{1}{2\tilde{\kappa}_l}\right).
\label{eqapph5}
\end{equation}

Figure~\ref{figapph1} shows the numerical results under the local or nonlocal measurement. Remarkably, we find that, when $0<\eta<1$, the scaling of the logarithmic negativity still exhibits the same transitions as in the case of the perfect measurements [cf. Figs.~\ref{fig2}(a), \ref{fig3}(a), and \ref{fig3}(b)]. Meanwhile, we also consider the case of $\eta=0$, where we take the ensemble average over the quantum trajectories. We show the time evolution of the logarithmic negativity in Fig.~\ref{figapph2}. We note that $\Gamma_\phi\left(t=0\right)=\left(1/2\right)1_{2N}$ is used as an initial state. We find that, when $\eta=0$, the logarithmic negativity suddenly vanishes in finite time.

%
%

%
\bibliography{MIPTFB}
\end{document}